\documentclass[onecolumn]{emulateapj}  

\usepackage{times}
\usepackage{verbatim}
\usepackage{graphicx,bm,amssymb}
\usepackage{epsfig}
\usepackage{amssymb}
\usepackage{natbib}
\usepackage{pstricks}
\usepackage{psfrag}
\usepackage[colorlinks=true,linkcolor=blue,citecolor=blue]{hyperref}
\usepackage{float}
\usepackage[caption = false]{subfig}
\usepackage{textcomp}
\usepackage{arydshln}
\usepackage{multirow}
\usepackage{lineno}

\newcommand{\be}{\begin{equation}}
\newcommand{\ee}{\end{equation}}
\newcommand{\bary}{\begin{eqnarray}}
\newcommand{\eary}{\end{eqnarray}}

\shorttitle{Off-axis GRBs}
\shortauthors{Fraija N.}

\begin{document}
\title{Modeling Gamma-ray burst Afterglow observations with an Off-axis Jet emission}
\author{N. Fraija$^{1\dagger}$, A. Galvan-Gamez$^{1}$, B. Betancourt Kamenetskaia$^{2,3}$,  M.G. Dainotti$^{4, 5}$,  S. Dichiara $^{6, 7}$, P. Veres$^{8}$,  R.~L.~Becerra$^{9}$ and A. C. Caligula do E. S. Pedreira$^{1}$}
\affil{$^1$ Instituto de Astronom\' ia, Universidad Nacional Aut\'onoma de M\'exico, Circuito Exterior, C.U., A. Postal 70-264, 04510 Cd. de M\'exico,  M\'exico.}
\affil{$^2$ LMU Physics Department, Ludwig Maxmillians University,Theresienstr. 37, 80333 Munich, Germany}
\affil{$^3$ TUM Physics Department, Technical University of Munich, James-Franck-Str, 85748 Garching, Germany}
\affil{$^4$ National Astronomical Observatory of Japan, Division of Science, Mitaka, 2-chome}
\affil{$^5$ Space Science Institute, Boulder, Colorado}
\affil{$^6$ Department of Astronomy, University of Maryland, College Park, MD 20742-4111, USA}
\affil{$^7$ Astrophysics Science Division, NASA Goddard Space Flight Center, 8800 Greenbelt Rd, Greenbelt, MD 20771, USA}
\affil{$^8$ Center for Space Plasma and Aeronomic Research (CSPAR), University of Alabama in Huntsville, Huntsville, AL 35899, USA}
\affil{$^9$ Instituto de Ciencias Nucleares, Universidad Nacional Aut\'onoma de M\'exico, Apartado Postal 70-264, 04510 M\'exico, CDMX, Mexico}

\email{$\dagger$nifraija@astro.unam.mx}

%

\date{\today} 
%
\begin{abstract}

Gamma-ray bursts (GRBs) are fascinating extragalactic objects. They represent a fantastic opportunity to investigate unique properties not exhibited in other sources. Multi-wavelength afterglow observations from some short- and long-duration GRBs reveal an atypical long-lasting emission that evolves differently from the canonical afterglow light curves favoring the off-axis emission. We present an analytical synchrotron afterglow scenario, and the hydrodynamical evolution of an off-axis top-hat jet decelerated in a stratified surrounding environment. The analytical synchrotron afterglow model is shown during the coasting, deceleration (off- and on-axis emission), and the post-jet-break decay phases, and the hydrodynamical evolution is computed by numerical simulations showing the time evolution of the Doppler factor, the half-opening angle, the bulk Lorentz factor, and the deceleration radius. We show that numerical simulations are in good agreement with those derived with our analytical approach. We apply the current synchrotron model and describe successfully the delayed non-thermal emission observed in a sample of long and short GRBs with evidence of off-axis emission. Furthermore, we provide constraints on the possible afterglow emission by requiring the multi-wavelength upper limits derived for the closest Swift-detected GRBs and promising gravitational-wave events. 
\end{abstract}


\section{Introduction}

Gamma-ray bursts (GRBs) are among the most powerful transient events in the Universe. These events are detected as brief, non-repeating flashes in the gamma-rays bands.  Depending on the burst duration (from milliseconds to thousands of seconds), GRBs are commonly classified as short (sGRBs) and long GRBs (lGRBs) \citep{kouveliotou1993}. A lGRB is associated with the core collapse (CC) of dying massive star \citep{1993ApJ...405..273W, 1998Natur.395..670G} that lead to supernova \citep[SNe;][]{1999Natur.401..453B, 2006ARA&A..44..507W}. At the same time, a sGRB is linked with the merger of a neutron star (NS) with a black hole \citep[BH,][]{1992ApJ...395L..83N} or two NSs \citep{1992ApJ...392L...9D, 1992Natur.357..472U, 1994MNRAS.270..480T, 2011MNRAS.413.2031M} as demonstrated by the historical gravitational-wave (GW) and electromagnetic detections of the GW170817 event \citep{PhysRevLett.119.161101}.\\

On August 17, 2017, the joint detection of two messengers of the fusion of two NSs was achieved for the first time: the GW event (GW170817) by the Laser Interferometer Gravitational-Wave Observatory (LIGO) and VIRGO \citep{PhysRevLett.119.161101,2017GCN.21520....1V} and the associated low-luminosity burst, GRB 170817A by the Fermi Gamma-ray Space Telescope \citep{2017GCN.21520....1V} and The INTErnational Gamma-Ray Astrophysics Laboratory \citep[INTEGRAL;][]{2017ApJ...848L..15S}. The results of this joint observation confirmed that NS fusion is a progenitor of sGRBs \citep{2017ApJ...848L..13A}.  The association of GW170817A with the near host galaxy NGC 4993, located at a redshift of $z\simeq0.01$ \citep{2017Sci...358.1556C,2017ApJ...848L..20M} suggested the presence of a local population of low-luminosity bursts following the merger of two NSs \citep{2017ApJ...848L..13A}.   Immediately,  GRB 170817A was followed up by large observational campaigns covering the X-ray \citep{troja2017a, 2018arXiv180103531M, 2018arXiv180502870A, 2018arXiv180106164D, 2017ATel11037....1M, 2018ATel11242....1H}, optical \citep{2018arXiv180102669L, 2018arXiv180103531M} and radio \citep[][]{2041-8205-848-2-L12, 2017arXiv171111573M, 2018ApJ...858L..15D, 2017Natur.547..425T} bands, among others. In order to describe the delayed multi-wavelength observations in timescales of days, synchrotron external-shock models radiated from off-axis top-hat jets \citep{troja2017a,2017ApJ...848L..20M, 2017arXiv171005905I, 2017ApJ...848L..21A, 2019ApJ...871..123F,2019ApJ...884...71F}, radially stratified ejecta \citep{2017arXiv171111573M, 2019ApJ...871..200F,2018ApJ...867...95H} and structured jets \citep{2017Sci...358.1559K, 2017arXiv171203237L} were proposed.   Later, analyses performed by Burns \textit{et al.}  \citep{2018ApJ...863L..34B} and other authors \citep{2018NatCo...9.4089T, 2018ApJ...863L..34B, 2016ApJ...833..151F} showed that GRB 150101B exhibits characteristics similar to GRB 170817A, in terms of its two-component structure and an undetected afterglow in a timescale of days followed by bright X-ray emission. Similar features in the multi-wavelength afterglow have been found in GRB 080503 \citep{2009ApJ...696.1871P},   GRB 140903A \citep{2016ApJ...827..102T} and GRB 160821B \citep{2017ApJ...835..181L, 2016GCN.19843....1S}.  On the other hand, several searches for afterglow emission around the closest bursts ($\lesssim 200\,{\rm Mpc}$) reported by the Burst Alert Telescope (BAT) instrument onboard the Neil Gehrels Swift Observatory \citep{2020MNRAS.492.5011D}  and the GW events by Advanced LIGO and Advanced VIRGO detectors \citep{2021PhRvX..11b1053A, 2021arXiv211103606T} have been performed without successful, but setting multi-wavelength upper limits.\\

The density profile of the medium surrounding a burst has been addressed previously in different contexts. For instance, for the case of SNe, \cite{1982ApJ...258..790C} studied the interaction of an adiabatic flow in a circumstellar density profile for Type II SNe of the form $\propto r^{-2}$. Ever since, subsequent authors have adopted this proposal for modeling the circumstellar medium, and it has even been applied in the research of different SN types. Examples of such studies are those by \cite{1996ApJ...472..257B}, \cite{2006ApJ...651.1005S}, \cite{2004MNRAS.354L..13K},  and \cite{1984ApJ...285L..63C}, among others.  Nevertheless, a generalization of this power law has also been considered. A later study, \cite{2012ApJ...747..118M} showed that the diversity might explain the spectral diversity of Type II luminous SNe in the density slope of the surrounding dense wind. To this effect, they proposed a wind density structure in the form $\propto r^{-k}$. They noticed that the ratio of the diffusion timescale in the optically thick region of the wind and the shock propagation timescale after the shock breakout strongly depends on the stratification parameter $k$, which led to differences in the spectral SN evolution.
On the other hand,  the requirement of a stratified environment condition has been proposed in some cases for modeling the multi-wavelength afterglows, such as in work by \cite{2013ApJ...776..120Y}. The authors analyzed more than one dozen GRBs and concluded that the circumburst environment could be neither a homogeneous nor a stellar-wind medium but something in between, with a general density distribution with a stratification parameter in the range $0.4 \leq k \leq 1.4$.   A more recent example of an analysis that links SN and GRB emission with a stratified environment is the one by \cite{2020A&A...639L..11I}, in which the authors studied SN 2020bvc. They found an excellent agreement with the GRB-associated, broad-lined Ic SN 1998bw; thus, it was categorized as a young broad-lined Ic SN. They also noted that its X-ray light curve was consistent with simulations of an off-axis GRB afterglow in a stratified medium with $k=1.5$; thus, this event represented the first case of an off-axis GRB discovered via its associated SN. It was later argued by \cite{2020ApJ...902...86H}, however, that such a model would predict an 8.5 GHz radio light curve several orders of magnitude more luminous than their measurements. Nevertheless, they stated that an off-axis jet could not be ruled out, and future radio observations would be needed.\\

In this work, we extend the analytical synchrotron afterglow scenario of the off-axis top-hat jet used to describe the multi-wavelength afterglow observations in GRB 170817A \cite[see][]{2019ApJ...884...71F} adding several ingredients. Here we consider i) the circumburst external medium as stratified with a profile density $\propto r^{-k}$ with ${\rm k}$ in the range of  $0\leq k < 3$, ii) the synchrotron radiation in self-absorption regime, iii) the afterglow emission during the transition from off-axis to on-axis before the lateral expansion phase (relativistic phase), iv) the hydrodynamical evolution computed by numerical simulations and v) a fraction of electrons accelerated by the shock front. We apply the current model to describe the delayed non-thermal emission observed in GRB 080503, GRB 140903A, GRB 150101B, GRB 160821B, and SN 2020bvc, and to provide constraints on the possible afterglow emission using multi-wavelength upper limits associated with the closest Swift-detected sGRBs and the promising GW events.  This paper is arranged as follows: Section 2 presents the analytical synchrotron scenario and an hydrodynamical evolution  of an off-axis top-hat jet decelerated in a stratified surrounding environment. In Section 3, we apply the proposed analytical model to describe the multi-wavelength observations of a sample of bursts and provide constraints to other ones. In Section 4, we present our conclusions.

\section{Off-axis top hat model}
\subsection{Radiative model}
Accelerated electrons in forward-shock models are distributed in accordance with their Lorentz factors ($\gamma_e$) and are described by the electron power index $p$ as $N(\gamma_e)\,d\gamma_e \propto \gamma_e^{-p}\,d\gamma_e$ for $\gamma_m\leq \gamma_{\rm e}$, where $\gamma_m=m_{\rm p}/m_{\rm e}g(p)\varepsilon_{\rm e}(\Gamma-1)\zeta^{-1}_e$ is the minimum electron Lorentz factor with $\Gamma$ is the bulk Lorentz factor, $m_{\rm p}$ and $m_{\rm e}$ the proton and electron mass, $\varepsilon_{\rm e}$ is the fraction of energy given to accelerate electrons, $\zeta_{e}$ denotes the fraction of electrons that were accelerated by the shock front \citep{2006MNRAS.369..197F} and $g(p)=\frac{p-2}{p-1}$. The comoving of the magnetic field strength in the blastwave $B'^2/(8\pi)=\varepsilon_Be$  is derived from the energy density $e=[(\hat\gamma\Gamma +1)/(\hat\gamma - 1)](\Gamma -1)n(r) m_pc^2$ with $\hat\gamma$ the adiabatic index \citep{1999MNRAS.309..513H} and its respective fraction given to magnetic field ($\varepsilon_B$). Hereafter, we adopt the unprimed and prime terms to refer them in the observer and  comoving frames, respectively. The term $\hat\gamma$ is the adiabatic index and $n(r) =A_{\rm k}\, r^{\rm -k}= \frac{\dot{M}_{\rm W}}{4\pi v_{\rm W}}\, r^{\rm -k}$, where $v_{\rm W}$ is the wind velocity and $\dot{M}_{\rm W}$ is the mass-loss rate. The sub-index ${\rm k}$ lies in the range $0\leq k \leq 3$, with ${\rm k=0}$ the constant-density medium ($A_{\rm 0}=n$), and ${\rm k = 2}$ the stellar wind ejected by its progenitor ($A_{\rm 2}\simeq A_{\rm W}\,3\times 10^{35}{\rm cm}^{-1}$) where $A_{\rm W}$ is the density parameter.  The cooling electron Lorentz factor is $\gamma_{\rm c}=(6\pi m_e c/\sigma_T)(1+Y)^{-1}\Gamma^{-1}B'^{-2}t^{-1}$, where $\sigma_T$ is the cross-section in the Thomson regime and $Y$ is the Compton parameter \citep{2001ApJ...548..787S, 2010ApJ...712.1232W}.  The synchrotron spectral breaks and the synchrotron radiation power per electron in the comoving frame are given by $\nu'_{\rm i}=q_e/(2\pi m_ec)\gamma^{2}_{\rm i}B'$ and $P'_{\nu'_m}\simeq \sqrt{3}q_e^3/(m_ec^2)B'$, respectively,  with hereafter the subindex ${\rm i=m}$ and ${\rm c}$ for the characteristic and cooling break, and the constants $q_e$ and $c$ the elementary charge and the speed of light, respectively \citep[e.g., see][]{1998ApJ...497L..17S, 2015ApJ...804..105F}. The synchrotron spectral breaks in the  self-absorption regime are derived from  $\nu'_{\rm a,1}=\nu'_{\rm c}\tau^{\frac35}_{0,m}$,  $\nu'_{\rm a,2}=\nu'_{\rm m}\tau^{\frac{2}{p+4}}_{0,m}$ and $\nu'_{\rm a,3}=\nu'_{\rm m}\tau^{\frac35}_{0,c}$ with the optical depth given by $\tau_{0,i}\simeq\frac{5}{3-k}\frac{q_en(r)r}{B'\gamma^5_{\rm i}}$ with $r$ the shock radius \citep{1998ApJ...501..772P}.   Considering  the total number of emitting electrons $N_e=(\Omega/4\pi)\, n(r) \frac{4\pi}{3-k} r^3$ and also taking into account the transformation laws for  the solid angle ($\Omega= \Omega'/\delta^2_D$), the radiation power ($P_{\nu_m}=\delta_D/(1+z) P'_{\nu'_m}$) and the spectral breaks  ($\nu_{\rm i}=\delta_D/(1+z)\nu'_{\rm i}$), the maximum flux given by synchrotron radiation is

\be
F_{\rm \nu, max}=\frac{(1+z)^2\delta^3_D}{4\pi d_z^2}N_eP'_{\nu'_m}\,,
\ee
 where {\small $d_{\rm z}=(1+z)\frac{c}{H_0}\int^z_0\,\frac{d\tilde{z}}{\sqrt{\Omega_{\rm M}(1+\tilde{z})^3+\Omega_\Lambda}}$}  \citep{1972gcpa.book.....W}  is the luminosiy distance, $r=\delta_D/(1+z) \Gamma\beta c t$ is the shock radius, and $\delta_D=\frac{1}{\Gamma(1-\mu\beta)}$ is the Doppler factor with $\mu=\cos \Delta \theta$, $\beta=v/c$ with $v$ the velocity of the material, and $\Delta \theta=\theta_{\rm obs} - \theta_{\rm j}$ is given by the viewing angle ($\theta_{\rm obs}$) and the half-opening angle of the jet ($\theta_{\rm j}$). We assume  for the cosmological constants a spatially flat universe $\Lambda$CDM model with  $H_0=69.6\,{\rm km\,s^{-1}\,Mpc^{-1}}$, $\Omega_{\rm M}=0.286$ and  $\Omega_\Lambda=0.714$ \citep{2016A&A...594A..13P}.

\subsection{Hydrodynamical evolution vs analytical approach}

\subsubsection{Hydrodynamical evolution}

We consider the dynamical equations proposed by \cite{1999MNRAS.309..513H, 2000MNRAS.316..943H}. The dynamical evolution of the relativistic outflow into the circumburst medium can be described  by

\bary
\frac{dr}{dt}&=&\beta c \Gamma \left(\Gamma + \sqrt{\Gamma^2-1}\right), \hspace{1cm}    \frac{dm}{dr}=2\pi\,(1-\cos\theta_{\rm j})r^2 n m_p\cr
\frac{d\theta_{\rm j}}{dt}&=&\frac{c_s}{r}(\Gamma + \sqrt{\Gamma^2-1}), \hspace{1.5cm} \frac{d\Gamma}{dm}=-\frac{\Gamma^2 -1}{M_{\rm ej}+\epsilon m + 2(1-\epsilon)\Gamma m}\,,
\eary

where $c_s=\sqrt{\hat\gamma(\gamma-1)(\Gamma-1)c^2/\left(1 + \hat\gamma(\Gamma -1)\right)}$ with $\hat\gamma\approx (4\Gamma + 1)/3\Gamma$,  $M_{\rm ej}$ is the initial value of the ejected mass and  $\epsilon$ is the radiative efficiency with $\epsilon=0$ in the adiabatic regime and $\epsilon=1$ in the fully radiative regime.  The previous equations are consistent with the self-similar solution during the ultra-relativistic \citep[Blandford-McKee solution;][]{1976PhFl...19.1130B} and the Newtonian phase (Sedov-Taylor solution), respectively, and consider the beaming effect \citep{1999ApJ...525..737R}. 

The observed quantities are integrated over the equal arrival time surface (EATS) determined by \citep{1997ApJ...491L..19W}

\be
t=(1+z)\int \frac{dr}{\beta\Gamma c\delta_D}\equiv {\rm const.}
\ee

\subsubsection{Analytical approach}

During the coasting phase (before the deceleration phase), the relativistic outflow is not affected by the circumburst medium, so the bulk Lorentz factor is constant $\Gamma_{\rm cp}= \Gamma_0$ and the radius evolves as $r=c\beta_0 t/[(1+z)(1-\beta_0\mu)]$ with $\beta_0=\sqrt{\Gamma^2_0-1}/\Gamma_0$. During the deceleration phase, the relativistic outflow transfers a large amount of its energy to the circumstellar medium driving a forward shock.   Considering the adiabatic evolution of the forward shock with an isotropic equivalent-kinetic energy  $E=\frac{4\pi}{3} m_pc^2 A_{\rm k} r^{3} \Gamma^2$\citep[Blandford-McKee solution;][]{1976PhFl...19.1130B} and a radial distance $r=c\beta_{\rm of} t/[(1+z)(1-\beta_{\rm of}\mu)]$, the bulk Lorentz factor evolves as 

\be\label{Gamma_dec_off}
\Gamma_{\rm of} = \left(\frac{3}{4\pi\,m_p c^{5-k}}\right)^{\frac12} \,(1+z)^{-\frac{k-3}{2}}  (1-\beta\cos\Delta\theta)^{-(k-3)}\,A_{\rm k}^{-\frac{1}{2}}\, E^{\frac{1}{2}}t^{\frac{k-3}{2}} \,,
\ee

with $\beta_{\rm of}=\sqrt{\Gamma^2_{\rm of}-1}/\Gamma_{\rm of}$. The deceleration time scale $t_{\rm dec}$ can be defined using eq. \ref{Gamma_dec_off}.   As the bulk Lorentz factor becomes $\Gamma\simeq1/\Delta\theta$, the observed flux becomes in our field of view. During the on-axis emission,  the bulk Lorentz factor in the adiabatic regime evolves as


\be\label{Gamma_dec_on}
\Gamma_{\rm on}= \left(\frac{3}{(2c)^{5-k}\pi\,m_p}\right)^{\frac{1}{8-2k}}\, (1+z)^{-\frac{k-3}{8-2k}} A_{\rm k}^{-\frac{1}{8-2k}} E^{\frac{1}{8-2k}}\,  t^{\frac{k-3}{8-2k}}\,,
\ee

with the radius $r\simeq 2\beta_{\rm on} \Gamma^2_{\rm on} c t/(1+z)$ and $\beta_{\rm on}=\sqrt{\Gamma^2_{\rm on}-1}/\Gamma_{\rm on}$, before the outflow enters to the post-jet-break decay phase ($\Gamma\simeq 1/\theta_j$).   During the post-jet-break decay phase, the bulk Lorentz factor evolves as 

\be\label{Gamma_dec_le}
\Gamma_{\rm le}= \left(\frac{3}{(2c)^{5-k}\pi\,m_p}\right)^{\frac{1}{6-2k}}\, (1+z)^{-\frac{k-3}{6-2k}} A_{\rm k}^{-\frac{1}{6-2k}} E^{\frac{1}{6-2k}} \theta_j^{\frac{1}{3-k}}  t^{\frac{k-3}{6-2k}}\,,
\ee

with the shock's radius and velocity given by $r\simeq 2\beta_{\rm le} \Gamma^2_{\rm le} c t/(1+z)$ and $\beta_{\rm le}=\sqrt{\Gamma^2_{\rm le}-1}/\Gamma_{\rm le}$, respectively.   We summarize the evolution of the bulk Lorentz factor as

{\small
\begin{eqnarray}
\label{gammas}
\Gamma \propto \cases{ 
t^0,\hspace{1.5cm}  t < t_{\rm dec} , \cr
t^{\frac{k-3}{2}},\hspace{1.1cm}  t_{\rm dec} \leq t\leq t_{\rm pk}, \cr
t^{\frac{k-3}{8-2k}}, \hspace{0.9cm} t_{\rm pk} \leq t\leq t_{\rm br} ,\hspace{.2cm} \cr
t^{\frac{k-3}{6-2k}}\, \hspace{1.2cm} t_{\rm br} \leq t,\hspace{.2cm}\cr
}
\end{eqnarray}
}

where the respective timescales are 

\bary\label{t_dec}
t_{\rm dec}&=&t_{\rm dec,0} (1+z) (1-\beta\cos\Delta\theta)^2 A_{\rm k}^{-\frac{1}{k-3}} E^{\frac{1}{k-3}}  \Gamma^{-\frac{2}{3-k}}\cr
t_{\rm pk}&=&t_{\rm pk,0} (1+z) A_{\rm k}^{\frac{1}{k-3}} E^{-\frac{1}{k-3}}  \Delta\theta^{-\frac{8-2k}{k-3}}\cr
t_{\rm br}&=&t_{\rm br,0} (1+z) A_{\rm k}^{-\frac{1}{3-k}} E^{\frac{1}{3-k}}  \theta_{\rm j}^{2}\,,
\eary

with {\small $t_{\rm dec,0}=(3/4\pi\,m_p c^{5-k})^{\frac{1}{3-k}}$} and {\small $t_{\rm pk,0}=t_{\rm br,0}=(3/(2c)^{5-k}\pi\,m_p )^{\frac{1}{3-k}}$}.  It is worth noting that the afterglow emission enters in the observer's field of view at $t_{\rm pk}$. For instance, the bulk Lorentz factor in a constant-density medium evolves as first $\propto t^0$ \citep{1999A&AS..138..537S}, then $t^{-\frac{3}{2}}$ \citep{2018arXiv180109712N, 2019ApJ...884...71F}, moreover $t^{-\frac{3}{8}}$ \citep{1998ApJ...497L..17S}, and finally $t^{-\frac{1}{2}}$ \citep{1999ApJ...519L..17S}, as expected.\\

The minimum and cooling electron Lorentz factors, the spectral breaks, the maximum flux and the synchrotron light curves during the coasting, deceleration (off- and on-axis) and post-jet-break decay phases are shown in Appendix \ref{Appendix} and Tables~\ref{Table2}, \ref{Table3} and \ref{Table1}.\\

\subsection{Comparison of our model with previous simulations}

\begin{figure}
{\centering
\resizebox*{0.9\textwidth}{0.5\textheight}
{\includegraphics{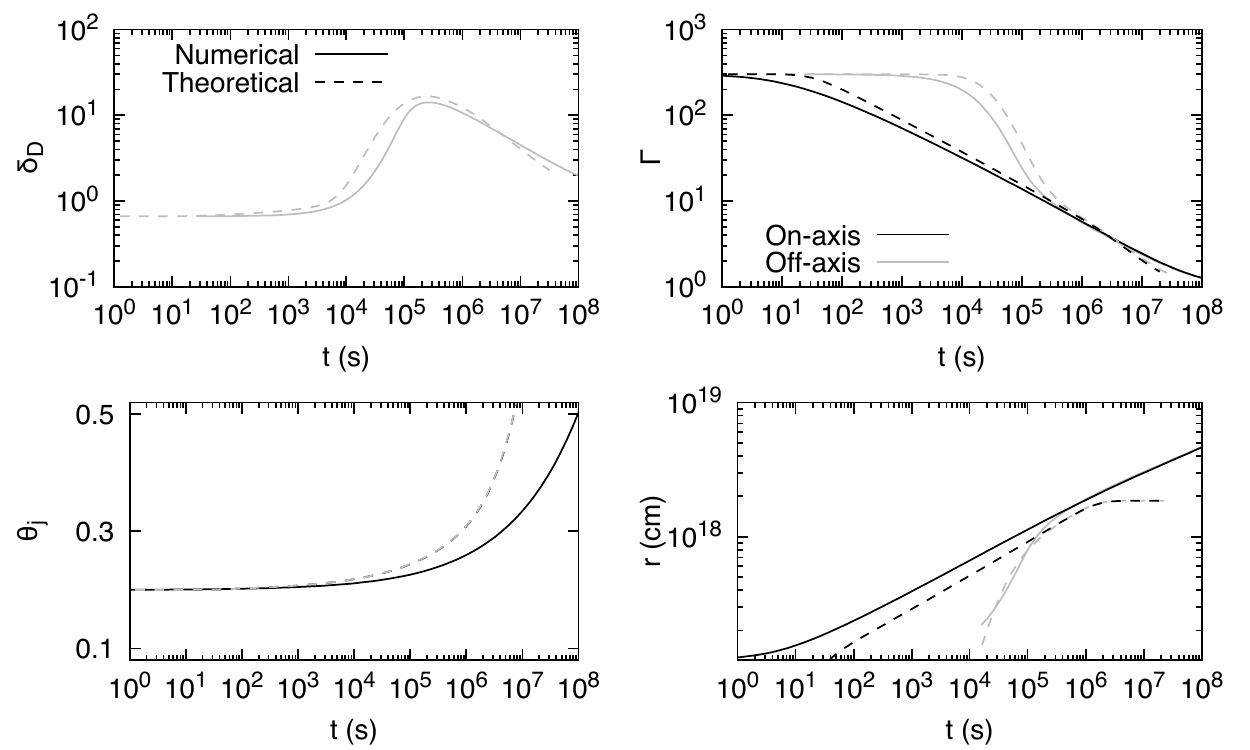}}}
\caption{Comparison of numerical simulations (solid lines) and a theoretical approach (dashed lines) of Doppler factor ($\delta_{\rm D}$), the jet's opening angle $\theta_{\rm j}$, the bulk Lorentz factor ($\Gamma_{\rm j}$) and the deceleration radius ($r$), all of them evolving in a constant circumburst medium. Variables observed for a observed off- and on-axis are in black and gray color, respectively. The following parameters $E=10^{54}\,{\rm erg}$, $n=1.0 \, {\rm cm^{-3}}$, $\Delta \theta=5.72 \, {\rm deg}$ and $\Gamma_0=300$ are used.}
\label{comparison}
\end{figure}

Figure~\ref{comparison} presents examples of the time evolution of the Doppler factor, the half-opening angle, the bulk Lorentz factor, and the deceleration radius, all in a constant circumburst medium ($k=0$). The solid lines characterize the numerical simulations, while the dashed lines stand-in for the theoretical approximations, which are detailed in Appendix \ref{Appendix}. Lines in black color corresponds to quantities observed off-axis and in gray color the ones observed on-axis.\\ 

For the case of $\delta_D$, the upper left panel shows an excellent agreement between the simulations and the model at early times up to $t_{\rm obs}\approx 10^{2}\, {\rm s}$. After this point, there are very slight variations between both solutions, specifically in the steepness of the rise and fall of this quantity. The numerical simulation predicts a sharper peak, while the theoretical approximation anticipates a wider profile. Despite this slight difference, both curves drop in parallel, showing that their delinquent behavior follows the same power law. In the case of the half-opening angle $\theta_{\rm j}$ (lower left panel), as in the case of $\delta_D$, the early time evolution is the same between both curves. The difference in their behavior is first made apparent at $t_{\rm obs}\approx10^{2}\, {\rm s}$. At this time, the theoretical solution presents a faster rise than the simulation. The initial variation between both curves keeps increasing and becomes substantial at times $\geq10^{6}\, {\rm s}$.

The upper right panel compares the bulk Lorentz factors when the emission is off- and on-axis. For the off-axis case, the same features mentioned in the last two panels are repeated, namely perfect agreement at early times and variations at late times. However, the on-axis emission is different as there are slight differences initially, but both solutions tend to be the same as time progresses. Finally, the right panel on the bottom presents the shock's radius for the on- and off-axis cases. For the on-axis curves, the simulation and the theoretical approximation differ early, but they unite late. There is also a difference in their shapes at the beginning, as the theoretical curve is a straight line, which corresponds to a single power law. At the same time, the simulation shows that there is a transition from a flatter curve to a steeper one. The case of the off-axis emission presents similar differences. In general, the simulation and the theoretical variables exhibited in Figure~\ref{comparison} are in good agreement.

\subsection{Analysis and description of the synchrotron light curves}

The analytical synchrotron afterglow model during the coasting, deceleration (off- and on-axis emission), and the post-jet-break decay phases are shown in Appendix~\ref{Appendix}. Table~\ref{Table2} shows the evolution of the synchrotron light curves, and Table~\ref{Table3} shows closure relations of synchrotron radiation as a function of $k$  during the coasting, deceleration (off- and on-axis emission), and the post-jet-break decay phases in the stratified environment. We can note in both tables that a break is expected around the transition time between fast- and slow-cooling regimes during the deceleration phase when the afterglow emission is seemed off-axis, but not during another phase.\\ 

\begin{figure}
\vspace{0.4cm}
{\centering
\resizebox*{0.9\textwidth}{0.5\textheight}
{\includegraphics{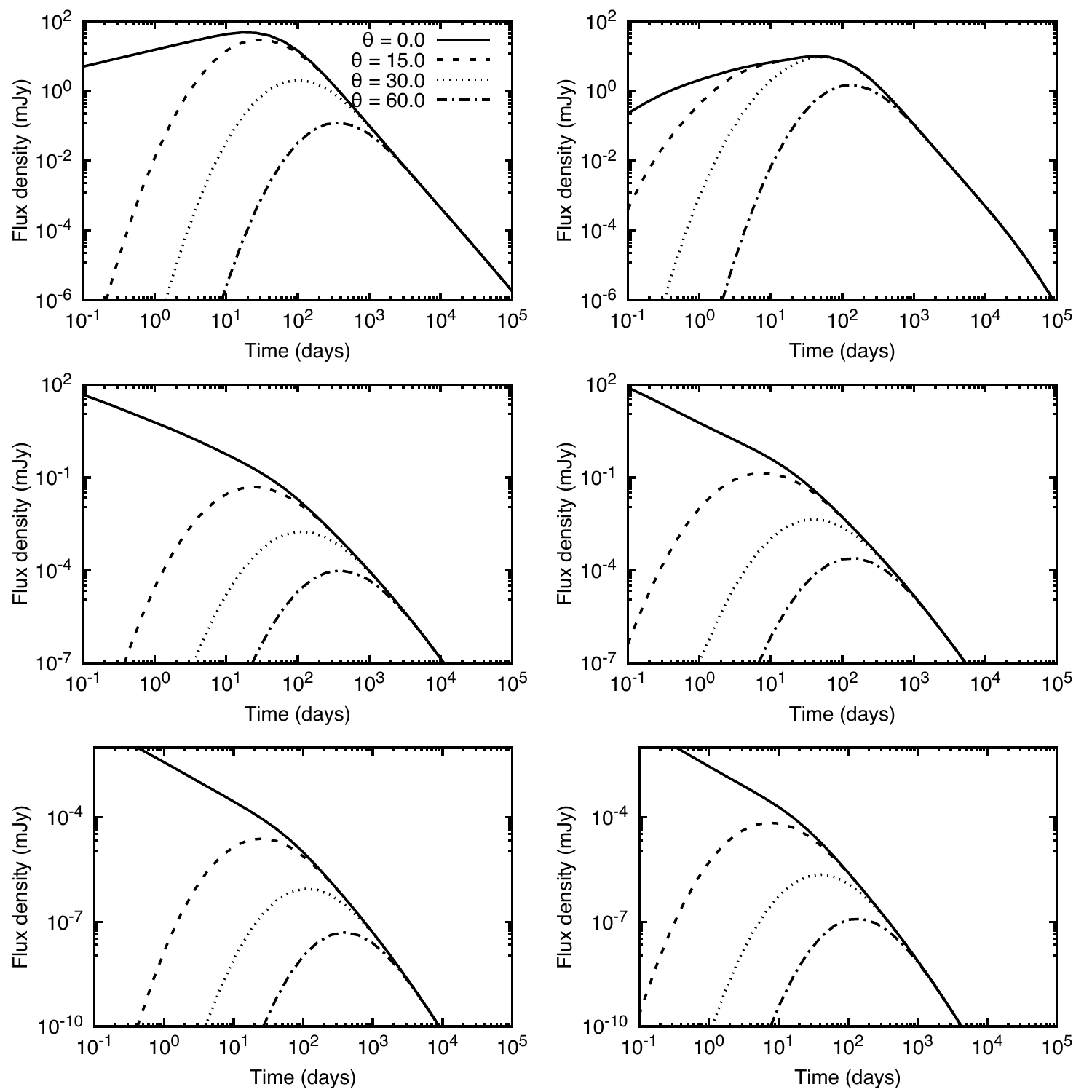}}}
\caption{Synchrotron light curves generated by the deceleration of the off-axis jet for $k = 0$ (left) and $k = 1$ (right). Panels from top to bottom correspond to radio (1.6 GHz), optical (R) and X-ray (1~keV) bands, respectively. The following parameters $E=10^{51}\,{\rm erg}$, $\varepsilon_{\rm B}=10^{-4}$, $\varepsilon_{\rm e}=10^{-1}$, $\Gamma=300$, $p=2.2$, $\zeta_e=1.0$ and $d_z=6.6\,{\rm Gpc}$ are used.}   
\label{k_0-k_1}
\end{figure}

\begin{figure}
\vspace{0.4cm}
{\centering
\resizebox*{0.9\textwidth}{0.5\textheight}
{\includegraphics{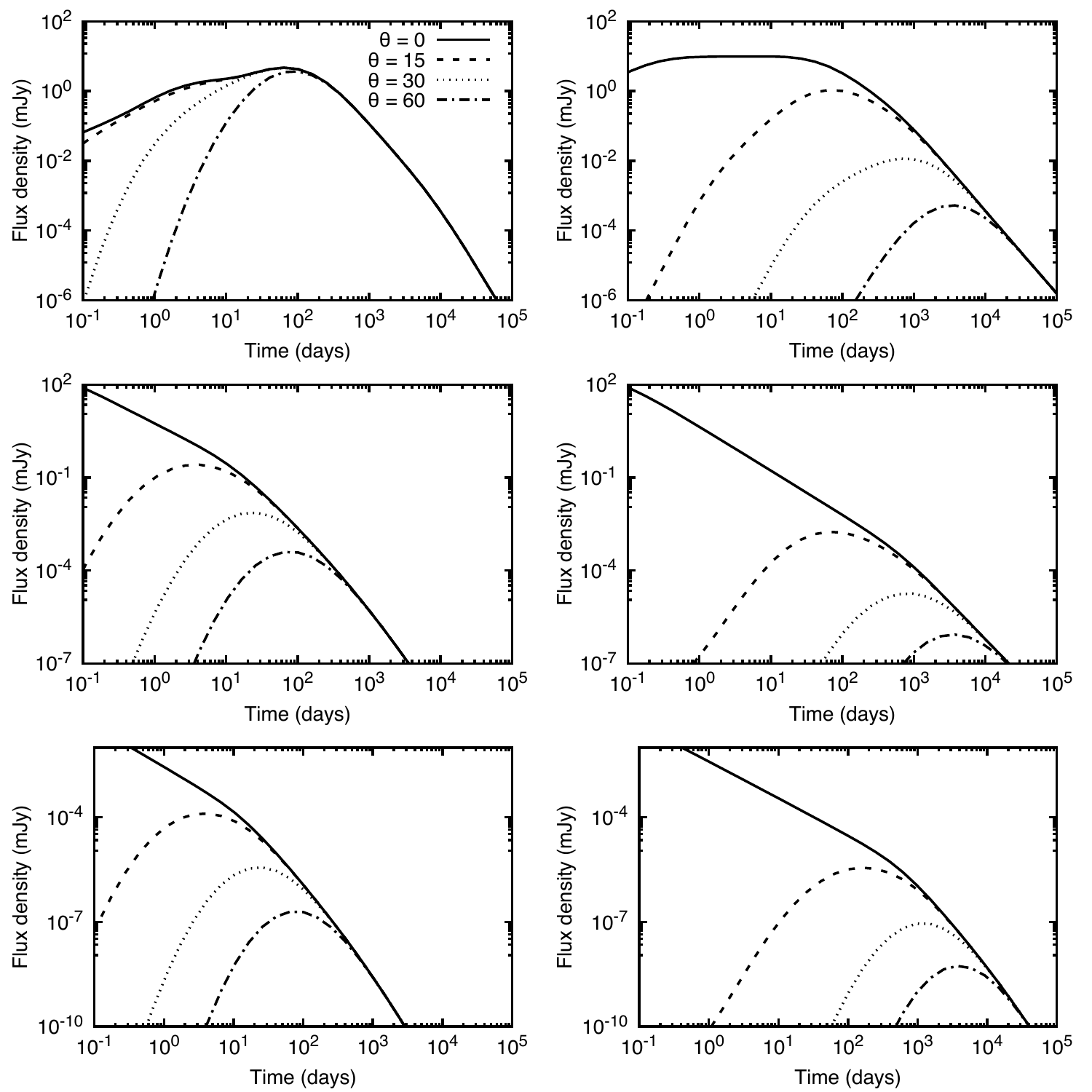}}}
\caption{Synchrotron light curves generated by the deceleration of the off-axis jet for $k = 1.5$ (left) and $k = 2$ (right). Panels from top to bottom correspond to radio (1.6 GHz), optical (R) and X-ray (1 keV) bands, respectively. The following parameters $E=10^{51}\,{\rm erg}$, $\varepsilon_{\rm B}=10^{-4}$, $\varepsilon_{\rm e}=10^{-1}$, $\Gamma=300$, $p=2.2$, $\zeta_e=1.0$ and $d_z=6.6\,{\rm Gpc}$ are used.}   
\label{k_15-2}
\end{figure}

Figures~\ref{k_0-k_1} and \ref{k_15-2} show the predicted synchrotron light curves produced by the deceleration of the off-axis top-hat jet in the circumburst medium described by a density profile with ${\rm k}=0$, $1.0$, $1.5$ and $2.0$, respectively. Panels from top to bottom correspond to radio wavelength at 6 GHz, optical at R-band and X-rays at 1~keV for typical values of GRB afterglow.\footnote{$E=10^{54}\,{\rm erg}$, $\varepsilon_{\rm B}=10^{-4}$, $\varepsilon_{\rm e}=10^{-1}$, $\zeta_e=1.0$ and  $d_z=1\,{\rm Gpc}$. Henceforth, we adopt the convention $Q_{\rm x}=Q/10^{\rm x}$ in cgs units for all variables except angles. For angles, we adopt the convention $Q_{\rm x}=Q/{\rm x}$ in degrees.} These figures display that regardless of the viewing angle, the X-ray, optical, and radio fluxes increase gradually, reach the maximum value, and finally, they begin to decrease. It can be observed that as the viewing angle increases, the maximum value moves to later times. For the chosen parameters, in most cases the maximum value lies in the range $(10^{-1} - 10^2)\,{\rm days}$ for $0\,\leq \theta_{\rm obs}\leq 60\,{\rm deg}$. This is different, however, when the medium is like a stellar wind, as the right column of Figure~\ref{k_15-2} shows that the maximum happens later, namely in the range $(10^{-1} - 10^4)\,{\rm days}$. Figures \ref{k_0-k_1} and \ref{k_15-2} show that the bump is less evident in the light curves with $1.0\leq k \leq 2.0$. Then, a clear rebrightening in a timescale from days to weeks with GW detection could be associated with the deceleration of the off-axis jet launched by a BNS or BH-NS merger. The synchrotron fluxes in all these panels lie in the slow-cooling regime, although for a different set of parameters with values (for instance, the equivalent kinetic energy $E = 10^{54}\,{\rm erg}$, the uniform-density medium $n\approx 1\,{\rm \,cm^{-3}}$, the equipartition parameters $\varepsilon_{\rm e}= 0.1$ and $\varepsilon_{\rm B}= 10^{-4}$), these would lie in the fast-cooling regime.\\

Table~\ref{TableDensityParameter} shows the evolution of the density parameter in each cooling condition of  the synchrotron afterglow model. For instance,  the synchrotron light curve as a function of the density parameter in the cooling condition $\nu_{\rm a,1} < \nu_{\rm m} < \nu < \nu_{\rm c} $  is given by $F_{\nu}:~\propto A_{\rm k}^{\alpha_{\rm k}}$, with $\alpha_{\rm k}={\frac{p+5}{4}}$, ${\frac{11-p}{4}}$, ${\frac{2}{4-k}}$ and ${\frac{3-p}{4(3-k)}}$, for the coasting, deceleration (off- and on-axis) and post-jet-break decay phases, respectively.  Any transition between a stratified environment and  density-constant medium could be detected during the post jet-break phase and  the deceleration phase when the afterglow emission (for this cooling condition) is seemed on-axis but not off-axis. Table~\ref{TableDensityParameter} shows that in general during the coasting phase and the deceleration phase when the afterglow emission is seemed off-axis cannot be detected a transition between different environments.   Table~\ref{TableDensityParameter} displays that synchrotron fluxes do not depend on the density parameter when these are observed at the lowest-energy (${\rm \nu < min \{\nu_{a,1}, \nu_{a,2}, \nu_{a,3} \} }$) and the highest-energy (${\rm max \{\nu_m, \nu_c \} < \nu }$) frequencies when the afterglow emission is seemed off- and on-axis, respectively. During the post jet-break phase, any transition between a stratified environment and  density-constant medium will be better observed in low-energy frequencies, such as radio wavelengths.\\

The uniform medium with constant density is expected for sGRBs associated with binary compact objects, and the stratified medium with non-constant density is expected for lGRBs related to massive stars with different evolution at the end of their lives \citep{2005ApJ...631..435R,  2006A&A...460..105V}. For instance, an external stratified medium with $0.4\leq k \leq 1.4$ was found by \cite{2013ApJ...776..120Y} after modeling the afterglow emission in a GRB example, and a density profile with $k>2$ was reported by \cite{2008Sci...321..376K} after studying the accretion of the stellar envelope by a BH as the possible origin of the plateau phase in X-ray light curves.

\cite{1999ApJ...524L..47G} analyzed the prompt gamma-ray emission in the BATSE\footnote{Burst and Transient Source Experiment} detected burst GRB 980923. The light curve exhibited a main prompt episode lasting $\sim 32\,{\rm s}$ followed by a smooth emission tail that lasted $\sim 390\,{\rm s}$. The authors found that the spectrum in the smooth tail evolved as the synchrotron cooling break  $t^{-0.52\pm0.12}$, concluding that the gamma-ray emission was associated with the afterglow evolution and also it had begun during the prompt gamma-ray episode. Afterward, spectra analyses of GRB tails were done to identify early afterglows \citep{2005ApJ...635L.133B, 2006MNRAS.369..311Y}. We could identify the off-axis synchrotron emission from an off-axis outflow analyzing the spectral break evolution. In this case, the spectral breaks of the synchrotron radiation generated from the deceleration of the off-axis jet in the relativistic phase evolve as a $\nu_{\rm m}\propto t^{-\frac{6-k}3}$ and $\nu_{\rm c}\propto t^{\frac{2+k}{2}}$, respectively. For instance, with $k=1$ the characteristic and cooling breaks evolve as $\nu_{\rm m}\propto t^{-\frac{5}{3}}$ and $\nu_{\rm c}\propto t^{\frac{3}{2}}$, respectively, which fully different to those breaks that evolve when the afterglow emission is on-axis (e.g., $\nu_{\rm m}\propto t^{-\frac{3}{2}}$ and $\nu_{\rm c}\propto t^{-\frac{1}{6}}$).\\

\section{A sample of some bursts with evidence of off-axis afterglow emission}

\subsection{Multi-wavelength Observations}

\paragraph{GRB 080503}  On 2008 May 3 at 12:26:13 UTC the {\itshape Swift}-BAT instrument detected and triggered on the short burst GRB 080503 \citep{2008GCNR..138....1M, 2008GCN..7677....1U}. The prompt episode was evaluated in the ($15 - 150$) keV energy range and reported with a duration of $(0.32\pm 0.07)\,{\rm s}$, while its observed flux was measured to be $(1.2\pm 0.2)\times 10^{-7}\,{\rm erg\,cm^{-2}\,s^{-1}}$.  The {\itshape Swift} X-ray Telescope (XRT) instrument and Chandra ACIS-S satellite also performed subsequent observations in the X-ray band \citep{2008GCN..7674....1G, 2009ApJ...696.1871P}. For the case of {\itshape Swift}-XRT, it took data on the burst in the timeframe from $\sim$ 82 s to 1 day after the initial {\itshape Swift}-BAT trigger. On the other hand, Chandra measured the burst throughout two observational campaigns, the first from 2008 May 07 (19:18:23 UTC) to 08 (04:09:59 UTC), during which an X-ray flux was detected, which coincided with the location of the optical afterglow. The second campaign took place from 2008 May 25 (18:11:36) to 26 (03:04:28), during which the X-ray source was monitored. There was a lack of detection, but constraining limits were provided.   Several efforts were also taken to monitor this burst in the optical energy range. Observations and upper limits were obtained with the {\itshape Swift} Ultra-Violet Optical Telescope (UVOT) instrument \citep[UVOT;][]{2008GCN..7675....1B}, the Hubble Space Telescope (HST) using the Wide-Field Planetary Camera (WFPC2) in the F606W, F450W and F814W bands \citep{2008GCN..7703....1B,  2008GCN..7749....1P, 2008GCN..7678....1P, 2008GCN..7679....1P}, and with the Keck-I telescope equipped with LRIS \citep{2008GCN..7666....1P}. The Gemini-N observatory also observed the burst using the Multi-Object Spectrograph (GMNOS) through the g, r, i and z optical bands and NIRI through the Ks band \citep{2008GCN..7695....1P,  2008GCN..7667....1B}. 
Regarding the radio energy band, the Karl G. Jansky Very Large Array (VLA) was used to observe GRB 080503 at a frequency of 8.46 GHz without any detection but providing a 3-sigma upper limit \citep{2008GCN..7684....1F}. 

\paragraph{GRB 140903A}  The {\itshape Swift}-BAT detected  GRB 140903A at 15:00:30 UT on 2014 September 14 \citep{2014GCN.16763....1C}.  This burst was located with coordinates R.A.= 238.036$^{\circ}$ and  Dec = +27.578$^{\circ}$ (J2000). The BAT light curve exhibited a peak with a duration of 0.45 s \citep{2014GCN.16765....1C}.  The XRT instrument began observing the position of this burst 59 s after the BAT trigger and monitored the X-ray afterglow  until this emission faded below the detector sensitivity threshold.   The MAXI-GSC observed the position reported by BAT 12 s after the trigger. Although no object was detected, upper limits were derived at the ($4 - 10$) keV energy range.  The Chandra satellite began observing the X-ray afterglow $\sim$ 2.7 s after the BAT trigger \citep{2014GCN.16813....1S}.  This burst was observed in the optical r-band  ($20.4\pm0.5$) mag by the 2-m telescope Faulkes Telescope North $\sim$ 15.6 hours after the BAT trigger \citep{2014GCN.16781....1D}. The Nordic Optical Telescope (NOT) observed the field of this burst, reporting an optical emission of  ($20.1\pm 0.5$) and ($16.1\pm 0.3$) mag in the $R$- and $H$-bands. \cite{2014GCN.16774....1C} identified a strong absorption doublet feature of the wavelengths in the range of  795.4 and 796.5~nm \citep{2014GCN.16774....1C}. These lines were associated  to NaID in absorption and H-beta  in emission at the usual redshift of $z=0.351$. Later, the detection of optical variability, together with a coincident radio detection \citep{2014GCN.16777....1F}, confirmed the host association of this redshift \citep{2014GCN.16785....1C}.

\paragraph{GRB 150101B} The {\itshape Swift}-BAT and {\itshape Fermi}-GBM detected GRB 150101B at 15:23:35  and 15:24:34.468 UT on 2015 January 01, respectively \citep{2015GCN.17267....1C,  2015GCN.17295....1S, 2018ApJ...863L..34B}.   Data analysis of  {\itshape Swift}-BAT  revealed a bright source, constraining the location at  R.A.=188.044$^{\circ}$ and Dec: -10.956$^{\circ}$ (J2000).  The $\gamma$-ray pulse in the ($15 - 150$) keV band consisted of a single pulse with duration and fluence of ($0.012\pm0.001$) s and $F_{\gamma}=(6.1\pm 2.2)\times 10^{-8}\,{\rm erg\, cm^{-2}}$, respectively \citep{2016ApJ...829....7L}. Recently, \cite{2018ApJ...863L..34B} presented a new analysis of fine timescales revealing a two-component structure; a short hard spike followed by a longer soft tail. The total duration of the prompt episode shown as a two-component structure was $0.08\pm0.93$ s and the fluence was $(1.2\pm0.1)\times 10^{-7}\,{\rm erg\,cm^{-2}}$ for the main peak and $(2.0\pm0.2)\times 10^{-8}\,{\rm erg\,cm^{-2}}$ for the soft tail.  The Chandra X-ray Observatory ACIS-S reported two observations, 7.94 and 39.68 days after the BAT trigger, with durations of $\sim$ 4.1 hours each one  \citep{2016ApJ...833..151F}.   Optical/near IR  observations and upper limits were collected with  Magellan/Baade using IMACS with r, g, i and z optical bands,  Very Large Telescope (VLT) equipped with the High Acuity Wide field K-band Imager I (HAWK-I) with the J, H, K and Y optical bands \citep{2015GCN.17288....1F,  2015GCN.17281....1L},   TNG using NICS with the J optical band \citep{2015GCN.17326....1D}, Gemini-S instrumented  with GMOS (r band) \citep{2015GCN.17333....1F}, UKIRT with the instrument WFCAM (J and K bands) and  HST using  the WFC3  with the F160W and F606 W bands \citep{2016ApJ...833..151F}.  Spectroscopic observations in the wavelength range $530-850\,{\rm nm}$ of 2MASX J12320498-1056010,  revealed several prominent absorption features that could associate GRB 150101B to an early-type host galaxy  located at $z=0.1343\pm 0.0030$ \citep{2015GCN.17281....1L,2016ApJ...833..151F}.\\

\paragraph{GRB 160821B}   The {\itshape Swift}-BAT and {\itshape Fermi} Gamma-Ray Monitor (GBM) Instrument triggered and located  GRB 160821B at 22:29:13  and 22:29:13.33 UT on 2016 August 21, respectively \citep{2016GCN.19833....1S, 2016GCN.19843....1S}. The {\itshape Swift}-BAT light curve in the energy range of ($15 - 150$) keV exhibited a single peak with duration and total fluence of ($0.48\pm 0.07$) s and $(1.1\pm0.1)\times 10^{-7}\,{\rm erg\,cm^{-2}}$, respectively \citep{2016GCN.19844....1P}. The {\itshape Fermi}-GBM light curve  in the energy range of ($8 - 1000$) keV showed a single peak (similar to the {\itshape Swift}-BAT profile) with duration and total fluence of $\sim1.2$ s and $(2.52\pm0.19)\times 10^{-6}\,{\rm erg\,cm^{-2}}$, respectively \citep{2017ApJ...835..181L}. They estimated that the isotropic energy released in gamma-rays was $E_{\rm iso} = (2.1\pm 0.2) \times 10^{50}$ erg.  {\itshape Swift}-XRT started detecting photons 57 s  after the trigger time \citep{2016GCN.19833....1S}.      {\itshape Swift}-UVOT began observing the field of GRB 160821B 76~s after the {\itshape Swift}-BAT trigger.   Although no photons in the optical band were detected during the first hours, constraining upper limits were placed \citep{2016GCN.19837....1E}. The William Herschel Telescope on La Palma, detected diverse emission lines in visible (including H-beta, [OIII] (4959/5007) and H-alpha), locating this burst with a redshift of $z= 0.16$ \citep{2016GCN.19846....1L}.

\paragraph{SN 2020bvc} On 2020 February 04 14:52:48 UTC SN 2020bvc was first detected by the ASAS-SN Brutus instrument using the g-Sloan filter with a reported location of R.A.=$14^{\textrm{h}}:33^{\textrm{m}}:57.024^{\textrm{s}}$ and Dec=+$40^{\circ}:14':36.85''$ (J2000). It was associated to the host galaxy UGC 09379, with a redshift of $z=0.025235$ \citep{TNSDR2020381}. A later report confirmed this association and redshift and based on the blue featureless continuum and the absolute magnitude at the discovery of -18.1 classified this event as a young core-collapse supernova \citep{TNSDR2020403}.  A later analysis \citep{TNSAN202037} of the spectrum obtained with the SPRAT spectrograph on the 2~m robotic Liverpool Telescope showed excellent agreement with the GRB-associated, broad-lined SN Ic 1998bw, thus it was categorized as a young broad-lined Ic SN. It was noticed to have an extremely fast rise by a steep decay in the two days following the first detection. It was later shown to rise towards a second peak. The decay temporal index reported by \cite{2020A&A...639L..11I} was $\alpha_{\rm dec}=1.35\pm0.9$. Twelve days later, on February 16, the Very Large Array (VLA) observed the position of SN 2020bvc and detected a point source with a flux density of $66\ \mu\textrm{Jy}$ in the X-band and a luminosity of $1.3\times10^{27}\,{\rm erg\,s^{-1}\,Hz^{-1}}$ \citep{TNSAN202042}. On the same day, a 10ks Chandra observation was obtained. The data was reduced, and the spectrum fitted with a power-law source model with a flux of approximately $10^{-14}\, {\rm erg\, cm^{-2}\,s^{-1}}$ \citep{TNSAN202045}.

\subsection{Analysis and Discussion}



\begin{figure}
\vspace{0.4cm}
{\centering
\resizebox*{1.1\textwidth}{0.8\textheight}
{\includegraphics[angle=-90]{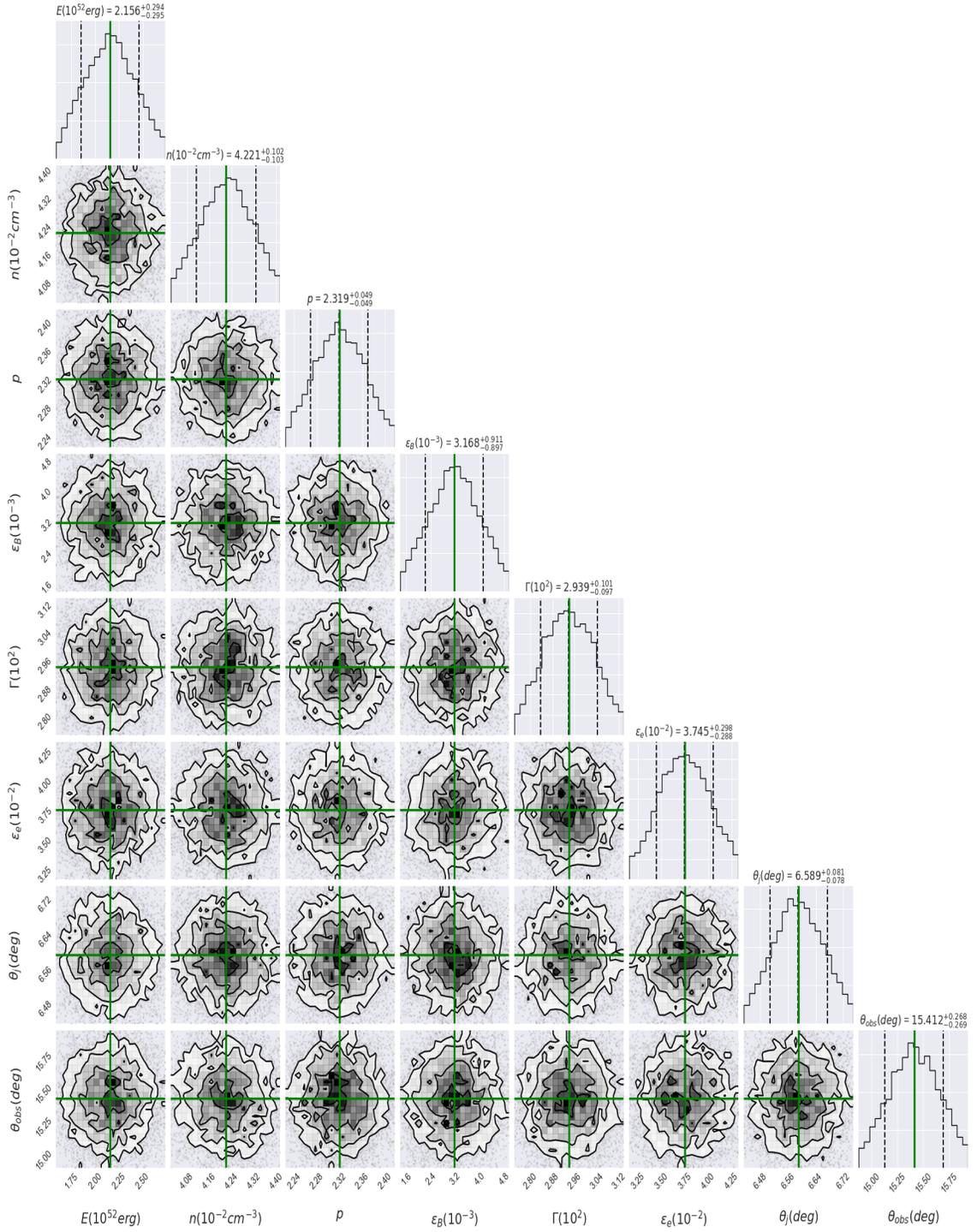}}}
\caption{Corner plot of the parameters derived from fitting the multi-wavelength light curves of GRB~080503 with a synchrotron off-axis model, together with median values (green lines) and 1$\sigma$ credible intervals (dashed lines). MCMC summary statistics for all parameters are listed in Table~\ref{par_mcmc}.   A set of normal distributions are made for the priors.}   
\label{GRB_080503_mcmc}
\end{figure}

\paragraph{GRB 080503}  We apply the Bayesian statistical approach of Markov-Chain Monte Carlo (MCMC) simulations to determine the best-fit values of the parameters that characterize the multi-wavelength afterglow observations with their upper limits \citep[e.g., see][]{2019ApJ...871..200F}.
A set of eight parameters, \{$E$, $n$,  $p$, $\varepsilon_{B}$, $\Gamma$, $\varepsilon_e$, $\theta_{\rm j}$ and $\theta_{\rm obs}$\}, is required by our synchrotron afterglow model evolving in a constant-density medium to describe the multi-wavelength observations.  A total of 17300 samples and 4100 tuning steps is used to describe the entire dataset. Normal distributions are used to characterize all the parameters. Corner plots illustrate the best-fit values and the median of the posterior distributions of the parameters, as shown in Figure~\ref{GRB_080503_mcmc}, respectively. In this figure, the best-fit values are highlighted in green, and the median of the posterior distributions are presented in Table~\ref{par_mcmc}.\\ 

\begin{figure}
\vspace{0.4cm}
{\centering
\resizebox*{0.9\textwidth}{0.5\textheight}
{\includegraphics{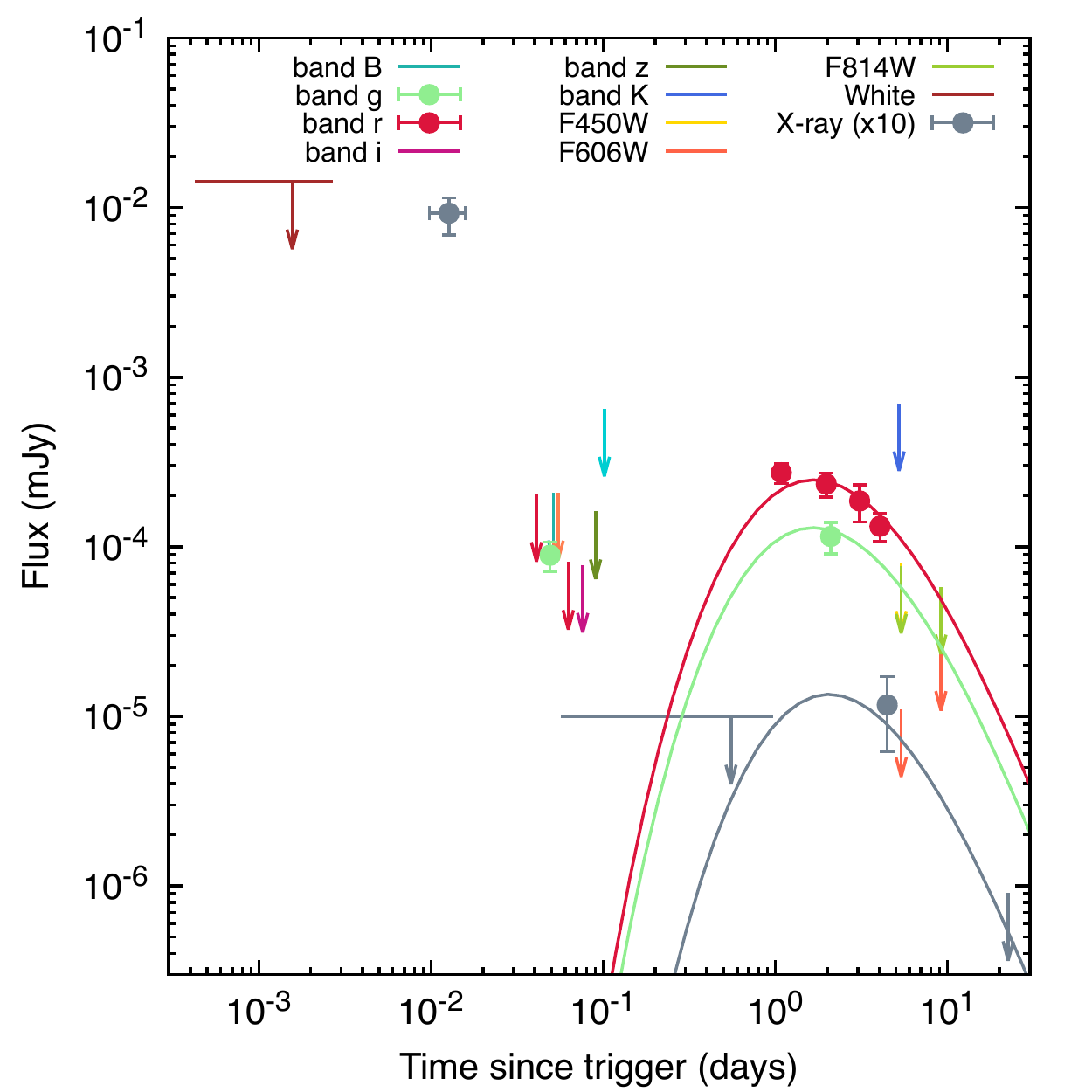}}}
\caption{X-ray and optical light curves of GRB 080503 with the best-fit curve of synchrotron afterglow model. The synchrotron light curves are shown at 1 keV (gray), R-band (red) and g-band (green).Data points are taken from \cite{2009ApJ...696.1871P}.}   
\label{GRB_080503}
\end{figure}

The multi-wavelength afterglow observations, together with the fits computed using the synchrotron off-axis model evolving in a constant-density medium are shown in Figure~\ref{GRB_080503}. The synchrotron light curves obtained with the same electron population and displayed in the optical (R and g) and  X-ray bands support the scenario of one-emitting zone inside an off-axis outflow.   The fact that the beaming cone of synchrotron radiation reaches our line of sight is compatible with the best-fit value of the viewing angle $\theta_{\rm obs}=15.412^{+0.268}_{-0.269}\,{\rm deg}$ and the re-brightening in all bands at $\sim$ one day. The best-fit value of the electron spectral index $p=2.319^{+0.049}_{-0.049}$ matches the typical value observed when forward-shocked electrons radiate by synchrotron emission \citep[see, e.g.][]{2015PhR...561....1K}. It confirms that these multi-wavelength observations originate during the afterglow.\\ 

The best-fit value of the constant-density medium $n=4.221^{+0.102}_{-0.103}\times 10^{-2}\,{\rm cm^{-3}}$ indicates that GRB 080503 took place in a medium with very low density comparable to an intergalactic density environment with $\sim 10^{-3}\,{\rm cm^{-3}}$. The best-fit values of  bulk Lorentz factor ($\Gamma=2.939^{+0.101}_{-0.078}\times 10^2$) and the equivalent-kinetic energy ($E=2.156^{+0.294}_{-0.295}\times 10^{52}\,{\rm erg}$)  indicate that synchrotron radiation is emitted from a narrowly collimated outflow.\\ 
%

\cite{2009ApJ...696.1871P} analyzed the multi-wavelength observations at $\sim$ one day. The authors dismissed the kilonova-like emission proposed by \cite{1998ApJ...507L..59L} and gave an afterglow explanation, pointing out that the X-ray and optical data had similar evolutions.  They hypothesized that the late optical and X-ray bumps might be explained in a slightly off-axis jet or a refreshed shock. The faint afterglow compared to the intense prompt emission could be described in the very low circumburst medium.  \cite{2015ApJ...807..163G} claimed that under certain requirements on the bulk Lorentz factor and the beaming angle of the relativistic jet, refreshed shocks in the synchrotron forward- and reverse-shock scenario could adequately describe the late re-brightening in GRB 080503.  Finally, \cite{2015ApJ...807..163G} proposed that the late optical peak was due to the emission from a magnetar-powered ``merger-nova", and the X-ray hump from magnetic dissipation of the magnetar dipole spin-down luminosity.   According to our analysis, these observations at $\sim$ one day are consistent with the afterglow emission found off-axis. It is worth noting that measurements of the linear polarization of the optical emission could discriminate between an on- and off-axis scenario and gravitational waves would provide information about the progenitor (a merger of NS - NS,  BH - NS or a stable NS.



\paragraph{GRB 140903A}

\begin{figure}
\vspace{0.4cm}
{\centering
\resizebox*{0.9\textwidth}{0.5\textheight}
{\includegraphics{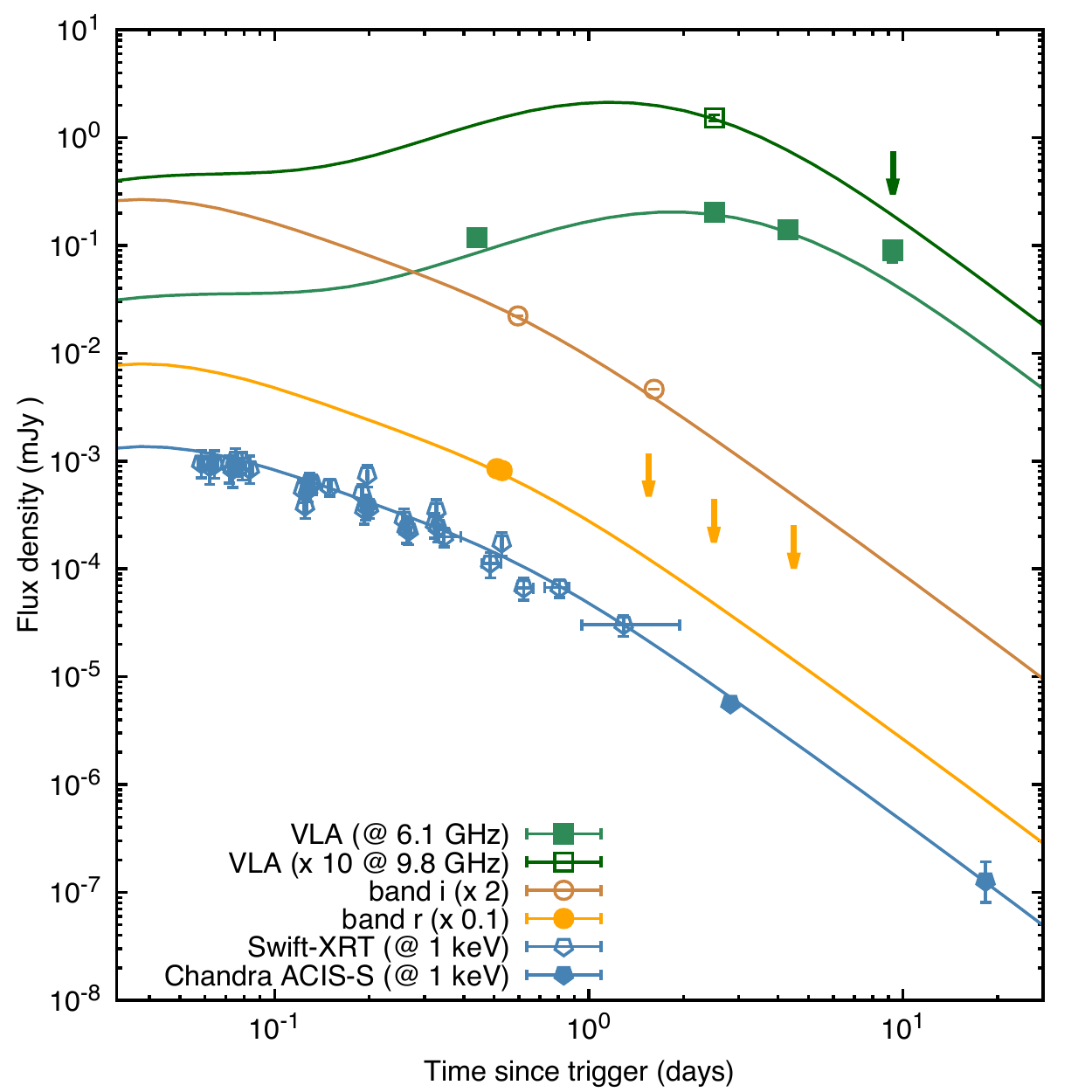}}}
\caption{X-ray, optical and radio light curves of GRB 140903A with the best-fit curve of synchrotron afterglow model. The synchrotron light curves are shown at 1 keV (blue), R-band (gold), i-band (brown) s, 9.8 GHz (dark green) and 6.1 GHz (light green). Data points are taken from \cite{2016ApJ...827..102T}.}   
\label{GRB_140903A}
\end{figure}

\begin{figure}
\vspace{0.4cm}
{\centering
\resizebox*{1.1\textwidth}{0.8\textheight}
{\includegraphics[angle=-90]{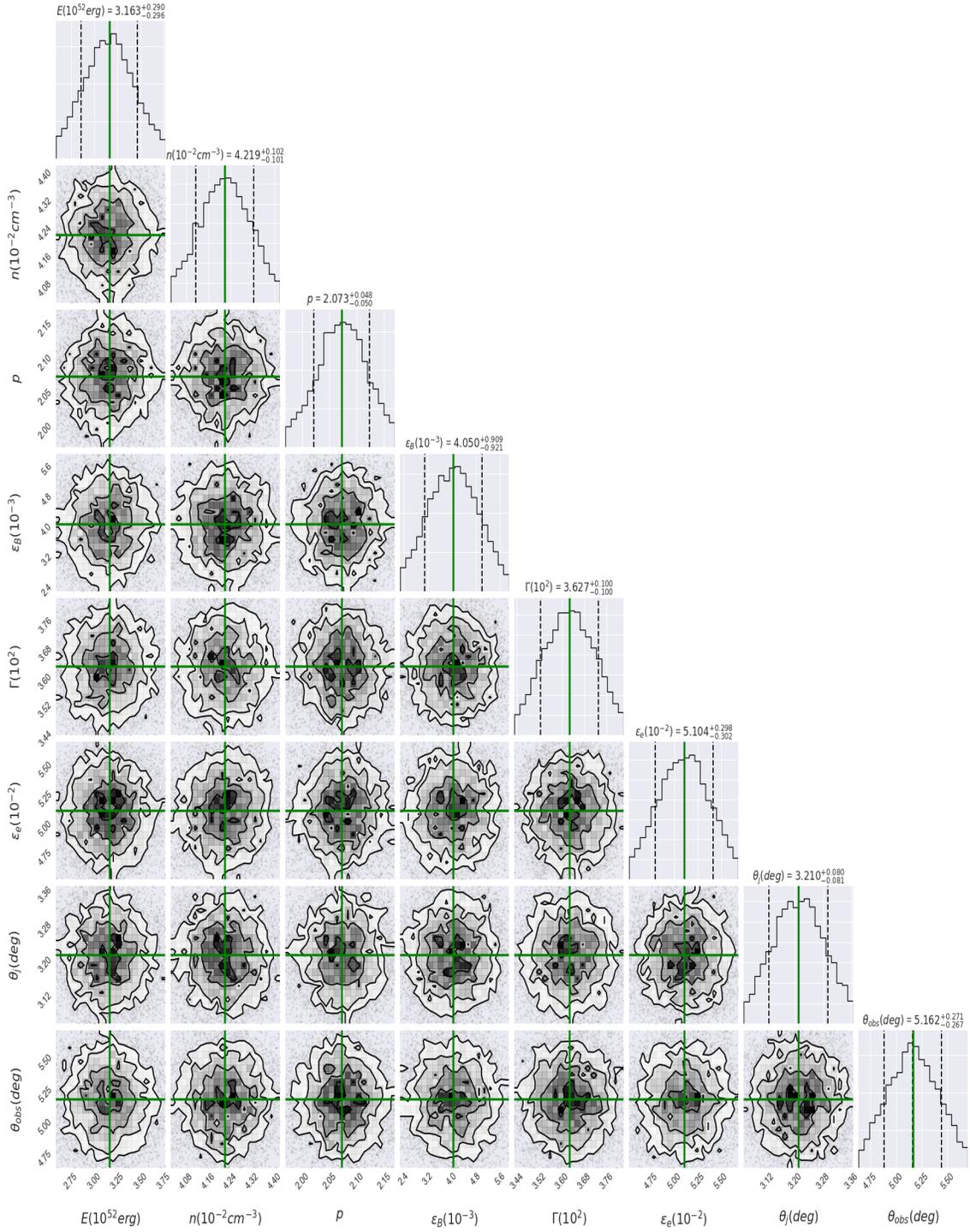}}}
\caption{The same as Figure~GRB 080503, but for GRB 140903A.}   
\label{GRB_140903A_mcmc}
\end{figure}

We once again conducted MCMC simulations with a set of eight parameters, \{$E$, $n$,  $p$, $\varepsilon_{B}$, $\Gamma$, $\varepsilon_e$, $\theta_{\rm j}$ and $\theta_{\rm obs}$\}, to find the best-fit values that describe the multi-wavelength afterglow observations with their upper limits. A set of eight parameters, \{$E$, $n$,  $p$, $\varepsilon_{B}$, $\Gamma$, $\varepsilon_e$, $\theta_{\rm j}$ and $\theta_{\rm obs}$\}, is required. To represent the entire observations in this scenario, a total of 16200 samples and 4150 tuning steps were used. Figure~\ref{GRB_140903A_mcmc} displays the best-fit values and the median of the posterior distributions of the parameters. In Table~\ref{par_mcmc}, the best-fit values are shown in green, and the median of the posterior distributions is presented.\\ 
The multi-wavelength observations of GRB 140903A are shown in Figure~\ref{GRB_140903A}, together with the fits derived using the synchrotron off-axis model evolving in a homogenous density.  The best-fit values of the viewing angle $\theta_{\rm obs}= 5.162^{+0.271}_{-0.267}\,{\rm deg}$ and the half-opening angle $\theta_{\rm j}= 3.210^{+0.080}_{-0.081}\,{\rm deg}$ indicate that the relativistic outflow was slightly off-axis.  The best-fit value of the constant density  medium $n=4.219^{+0.102}_{-0.101}\times 10^{-2}\,{\rm cm^{-3}}$ indicates that GRB 140903A took place in a medium  with very low density comparable to an intergalactic density environment. The best-fit values of  bulk Lorentz factor ($\Gamma=3.627^{+0.100}_{-0.100}\times 10^2$) and the equivalent-kinetic energies ($E=3.163^{+0.290}_{-0.296}\times 10^{52}\,{\rm erg}$)  indicate that synchrotron emission is radiated from a narrowly collimated jet. The best-fit values of the microphysical parameters $\varepsilon_{\rm e}= 5.104^{+0.298}_{-0.302}\times10^{-2}$ and $\varepsilon_{\rm B}= 4.050^{+0.909}_{-0.921}\times10^{-3}$, and the best-fit value of the spectral index of the shocked electrons $2.073^{+0.048}_{-0.050}$, are similar to those reported in \cite{2016ApJ...827..102T}. This spectral index is in the typical range to those accelerated in forward shocks \citep[see, e.g.][]{2015PhR...561....1K}, thus reaffirming the afterglow as its origin.\\

\cite{2016ApJ...827..102T} reported and analyzed the afterglow observations of GRB 140903A for the first two weeks.
Applying the fireball scenario, the authors demonstrated that this burst was caused by a collimated jet seen off-axis and was also connected with a compact binary object. The X-ray "plateau" seen in GRB 140903A was attributed to the energy injection into the decelerating blast wave by \cite{2017ApJ...835...73Z}. The authors then modelled the late afterglow emission, which required a half-opening angle of $\approx 3\,{\rm deg}$, similar to the value found with our model.  GRB 140903A was formed in a collimated outflow observed off-axis that decelerates in a uniform density, according to our findings.\\

\paragraph{GRB 150101B} 

\begin{figure}
\vspace{0.4cm}
{\centering
\resizebox*{1.1\textwidth}{0.8\textheight}
{\includegraphics[angle=-90]{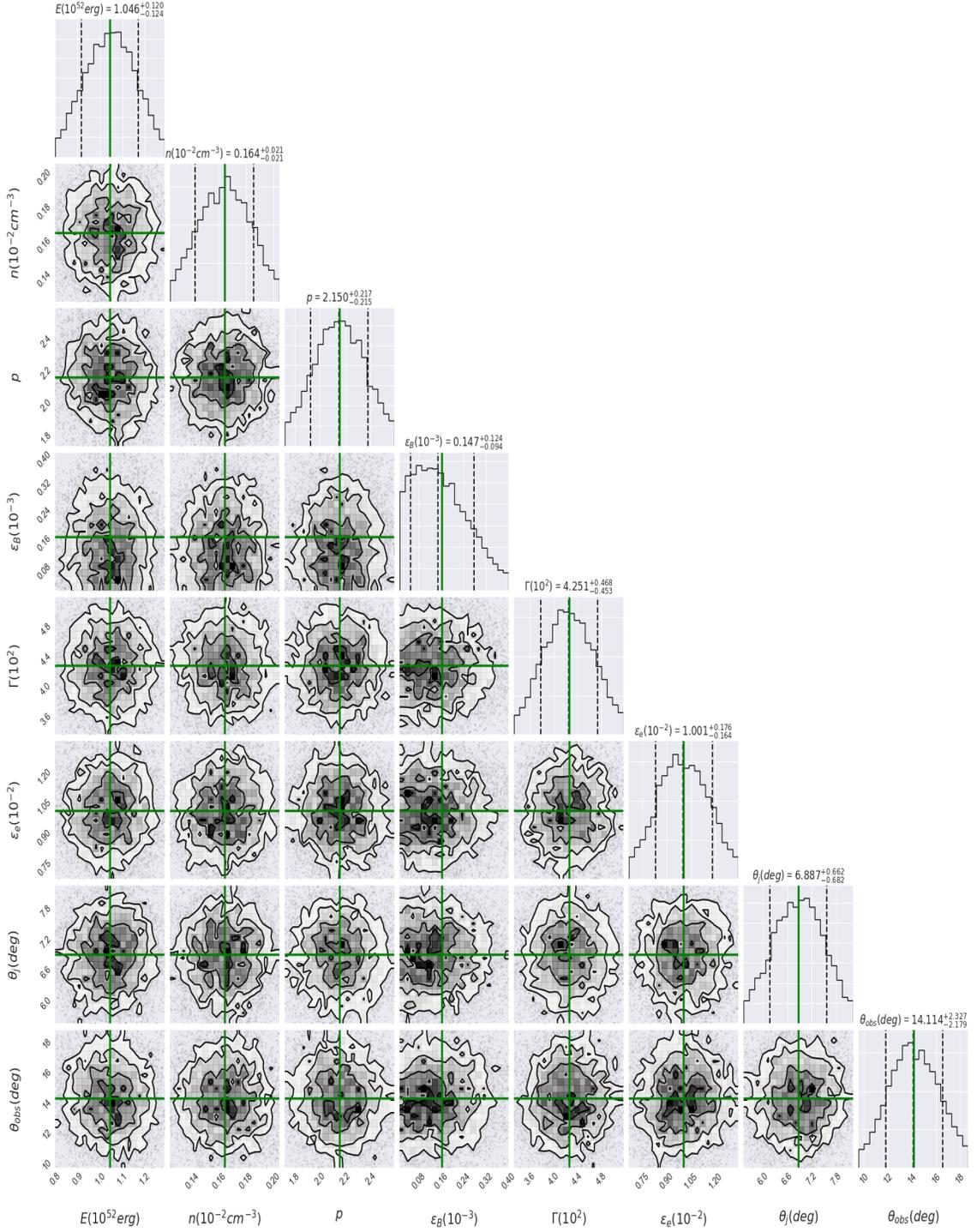}}}
\caption{The same as Figure~GRB 080503, but for GRB 150101B.}   
\label{GRB_150101B_mcmc}
\end{figure}

We use MCMC simulations with the eight parameters used for GRB 080503 and GRB 140903A to find the best-fit values that model the X-ray afterglow observations with the optical upper limits. To represent the entire data in this case, a total of 15900 samples and 4400 tuning steps are used. Figure~\ref{GRB_150101B_mcmc} displays the best-fit values and the median of the posterior distributions of the parameters. In Table~\ref{par_mcmc}, the best-fit values are shown in green, and the median of the posterior distributions is presented.\\ 

\begin{figure}
\vspace{0.4cm}
{\centering
\resizebox*{0.9\textwidth}{0.35\textheight}
{\includegraphics{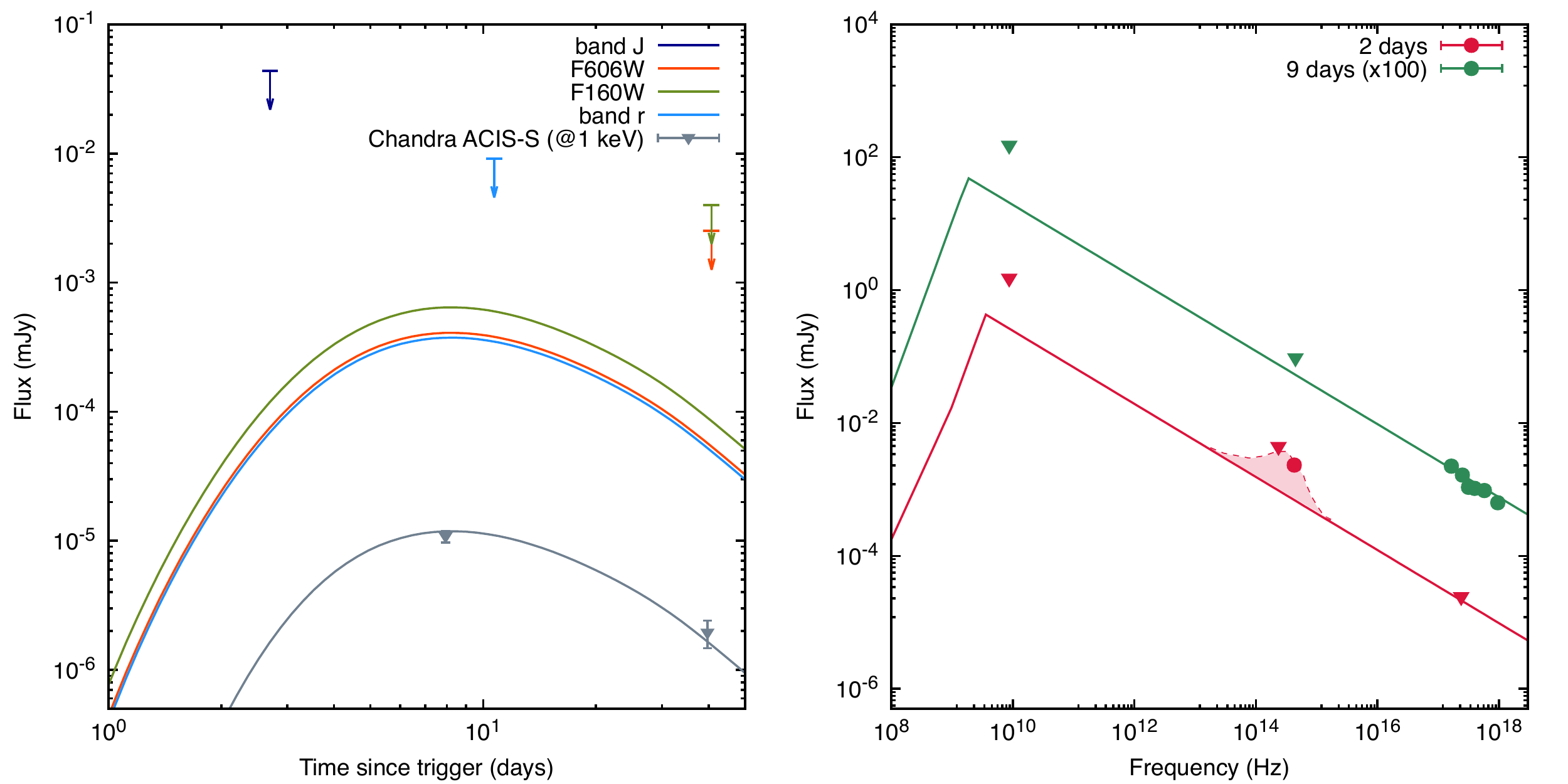}}}
\caption{Left: X-ray light curves with the optical upper limits of GRB 150101B with the best-fit curve of synchrotron afterglow model. The synchrotron light curves are shown at 1 keV (gray), R-band (blue), F606W filter (orange) and F160W filter (dark green). Data points are taken from \cite{2016ApJ...833..151F} and \cite{2018NatCo...9.4089T}.
Right: The broadband SEDs of the X-ray, optical and radio afterglow observations and upper limits  with the best-fit synchrotron curves (lines) at 2 days (red) and 9 (green) days, respectively. The red area corresponds to the spectrum of AT2017gfo, which is adapted by  \cite{2018NatCo...9.4089T}. 
}   
\label{GRB_150101B}
\end{figure}

The X-ray, optical and radio observations and upper limits, as well as the fit obtained using the synchrotron off-axis model evolving in  homogeneous density are shown in Figure~\ref{GRB_150101B}. The left-hand panel shows the light curves at 1 keV (gray), R-band (blue), F606W filter (orange) and F160W filter (dark green), and the right-hand panel displays the broadband SEDs at 2 (red) and 9 (green) days.  The red area corresponds to the spectrum of AT2017gfo, which is adapted by  \cite{2018NatCo...9.4089T}.   The best-fit values of the viewing angle $\theta_{\rm obs}= 14.114^{+2.327}_{-2.179}\,{\rm deg}$ and the half-opening angle $\theta_{\rm j}= 6.887^{+0.662}_{-0.682}\,{\rm deg}$ can explain the lack of X-ray emission during the first day.  The best-fit value of the constant density  medium $n=0.164^{+0.021}_{-0.021}\times 10^{-2}\,{\rm cm^{-3}}$ indicates that GRB 150101B, like other short bursts,  happened in an environment  with very low density. The best-fit values of  bulk Lorentz factor ($\Gamma=4.251^{+0.468}_{-0.453}\times 10^2$) and the equivalent-kinetic energy ($E=1.046^{+0.120}_{-0.124}\times 10^{52}\,{\rm erg}$) suggest that synchrotron afterglow emission is emitted  from a narrowly collimated jet. The values of the spectral index of the electron population, the circumburst density, the microphysical parameters, and the viewing angle disfavor the isotropic cocoon model reported in \cite{2018NatCo...9.4089T} and are consistent with the values of an outflow when homogeneous density is taken into account.    The best-fit values of the spectral index of the shocked electrons $2.150^{+0.217}_{-0.215}$ is similar to those reported in synchrotron afterglow models \citep[see, e.g.][]{2015PhR...561....1K}.\\

\paragraph{GRB 160821B}

\begin{figure}
\vspace{0.4cm}
{\centering
\resizebox*{1.1\textwidth}{0.8\textheight}
{\includegraphics[angle=-90]{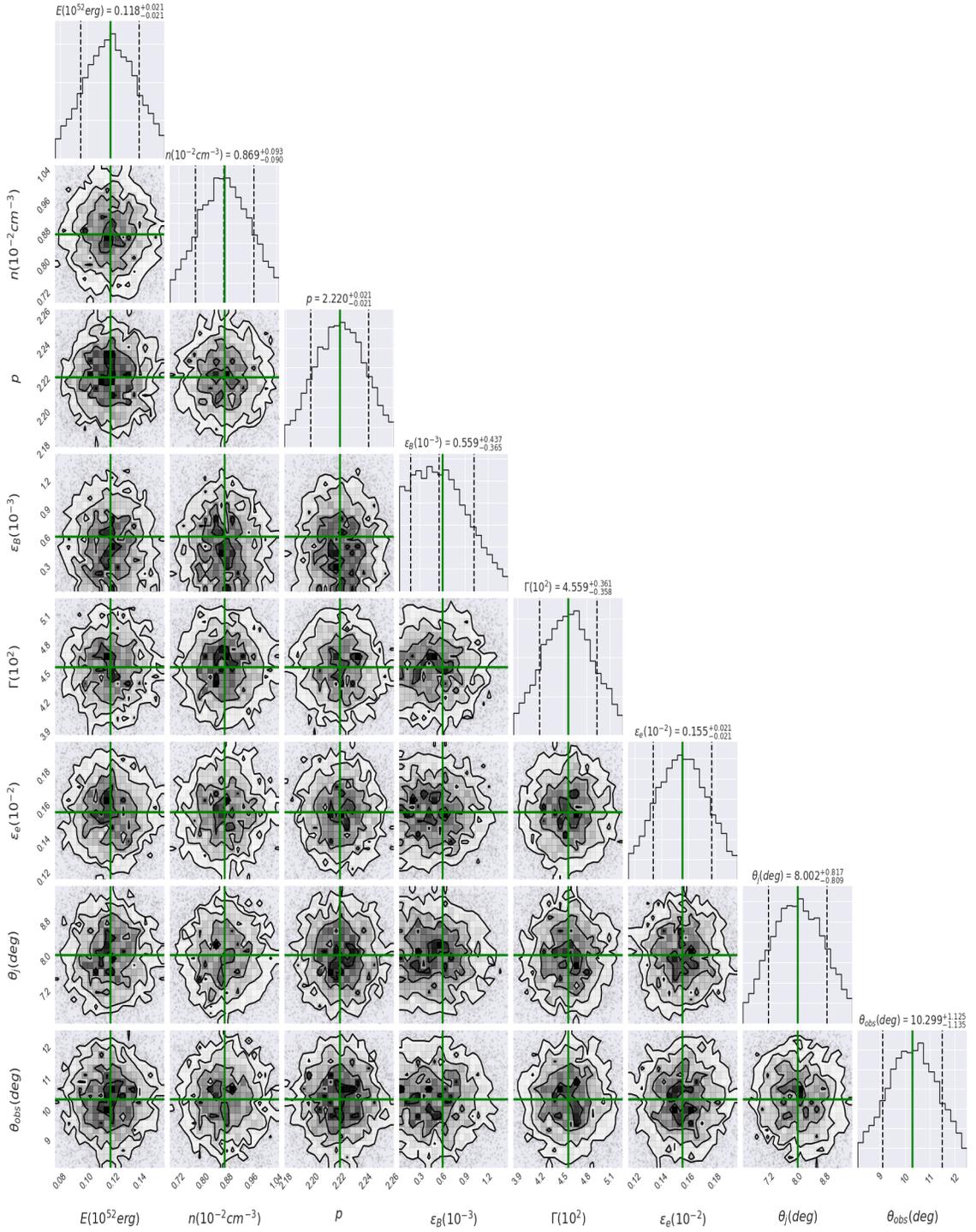}}}
\caption{The same as Figure~GRB 080503, but for GRB 160821B.}   
\label{GRB_160821B_mcmc}
\end{figure}

We require MCMC simulations with eight parameters used in the previous bursts to find the best-fit values that describe the multi-wavelength afterglow observations with their upper limits. To characterize the complete data in this scenario, a total of 17300 samples and 7400 tuning steps are used. Figure~\ref{GRB_160821B_mcmc} exhibits the best-fit values and the median of the posterior distributions of the parameters. In Table~\ref{par_mcmc}, the best-fit values and the medians of the posterior distributions are presented.\\

\begin{figure}
\vspace{0.4cm}
{\centering
\resizebox*{0.9\textwidth}{0.35\textheight}
{\includegraphics{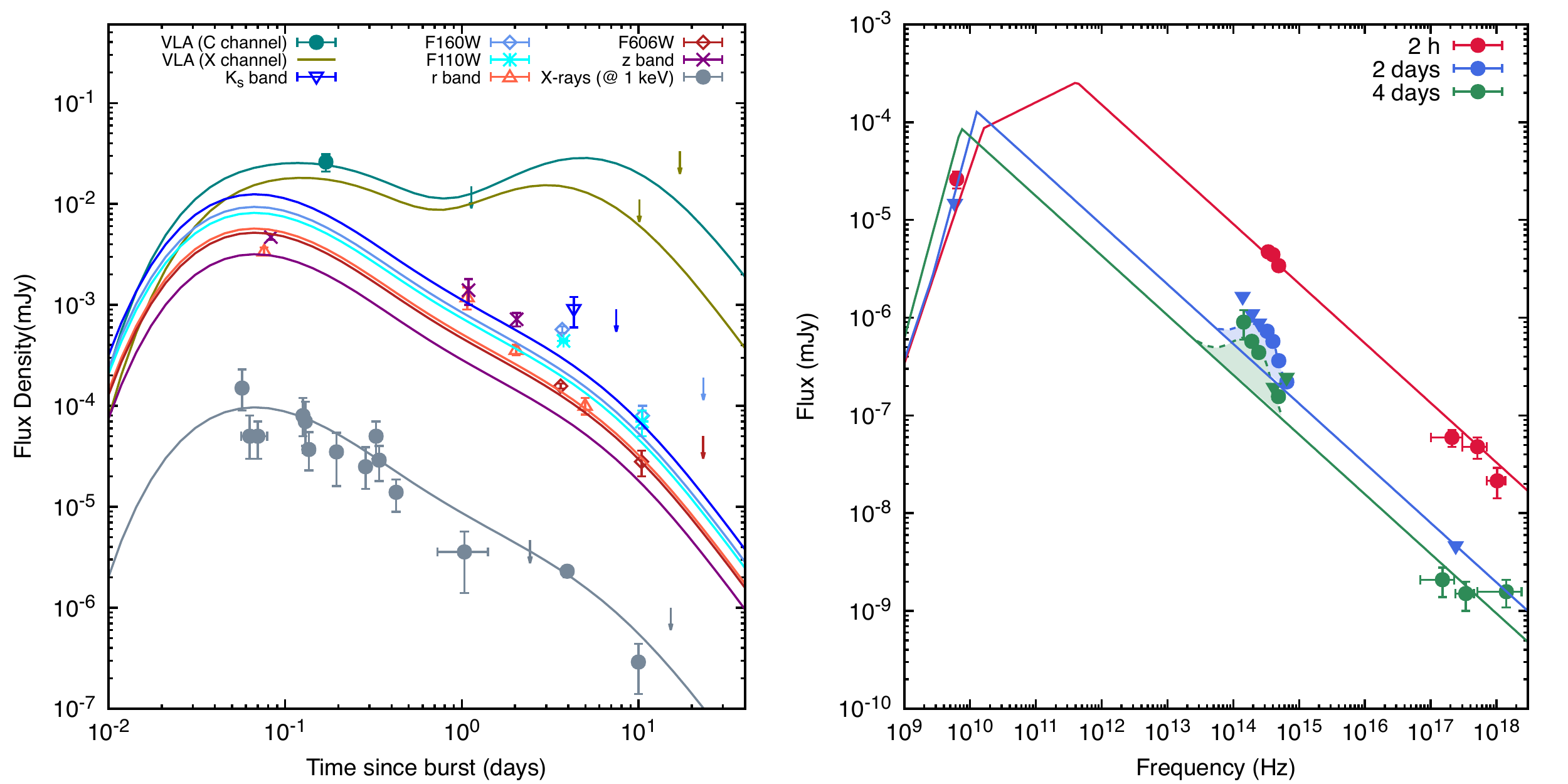}}}
\caption{Left: X-ray, optical and radio light curves of GRB 160821B with the best-fit curve of synchrotron afterglow model. The synchrotron light curves are shown at 1 keV (gray), z-band (purple), F606W filter (dark red), R-band (salmon), F110W filter (cyan), F160W filter (blue sky), K$_s$-band (blue), X-channel (olive) and C-channel (emerald green). Data points are taken from \cite{2019MNRAS.489.2104T}. Right: The broadband SEDs of the X-ray, optical and radio afterglow observations with the best-fit synchrotron curves (lines) at 2 h (red), 2 days (blue) and 4 (green) days, respectively. The shaded areas correspond to blackbody spectra with decreasing temperatures from \cite{2019MNRAS.489.2104T}.}   
\label{GRB_160821B}
\end{figure}

The multi-wavelength observations since 0.2 days after the GBM trigger are shown in Figure~\ref{GRB_160821B}, together with the fits found requiring the synchrotron off-axis model evolving in a homogeneous density. The left-hand panel exhibits the light curves at 1 keV (gray), z-band (purple), F606W filter (dark red), R-band (salmon), F110W filter (cyan), F160W filter (blue sky), K$_s$-band (blue), X-channel (olive) and C-channel (emerald green), and the right-hand panel shows the broadband SEDs of the X-ray, optical and radio afterglow observations at 2 h (red), 2 days (blue) and 4 days (green). The shaded areas in blue and green correspond to blackbody spectra with decreasing temperatures firstly suggested in \cite{2018ApJ...857..128J} and then confirmed by \cite{2019ApJ...883...48L} and \cite{2019MNRAS.489.2104T}. The kilonova emission is the most natural explanation for the ``new" radiation component.  The {\itshape Swift}-XRT-UVOT data were received from the public database from the official Swift website.\footnote{https://swift.gsfc.nasa.gov/cgi-bin/sdc/ql?} The C-band displays radio data, the white, v, b, u, UVW1, UVW2, and UVM2  bands display {\itshape Swift}-UVOT data, and the 1~keV band displays XRT data. Using the conversion factor proposed in \cite{2010A&A...519A.102E}, the flux density of XRT data is extrapolated from 10~keV to 1~keV.  The best-fit values of circumburst density $n=0.869^{+0.093}_{-0.090}\times 10^{-2}\,{\rm cm^{-3}}$, spectral index of shocked electrons $p=2.220^{+0.021}_{-0.021}$, viewing angle $\theta_{\rm obs}= 10.299^{+1.125}_{-1.135}\,{\rm deg}$ and the half-opening angle $\theta_{\rm j}= 8.002^{+0.817}_{-0.809}\,{\rm deg}$ are similar to those previously reported in \cite{2019arXiv190501290T}.   The best-fit values of the viewing angle $\theta_{\rm obs}= 10.299^{+1.125}_{-1.135}\,{\rm deg}$ and the half-opening angle $\theta_{\rm j}= 8.002^{+0.817}_{-0.809}\,{\rm deg}$ suggest that the relativistic outflow was slightly off-axis.
The best-fit values of  bulk Lorentz factor ($\Gamma=4.559^{+0.361}_{-0.358}\times 10^2$) and the equivalent-kinetic energy ($E=0.118^{+0.021}_{-0.021}\times 10^{52}\,{\rm erg}$) indicate that synchrotron afterglow emission is emitted  from a narrowly collimated jet.

\paragraph{SN 2020bvc}

\begin{figure}
{\centering
{\includegraphics{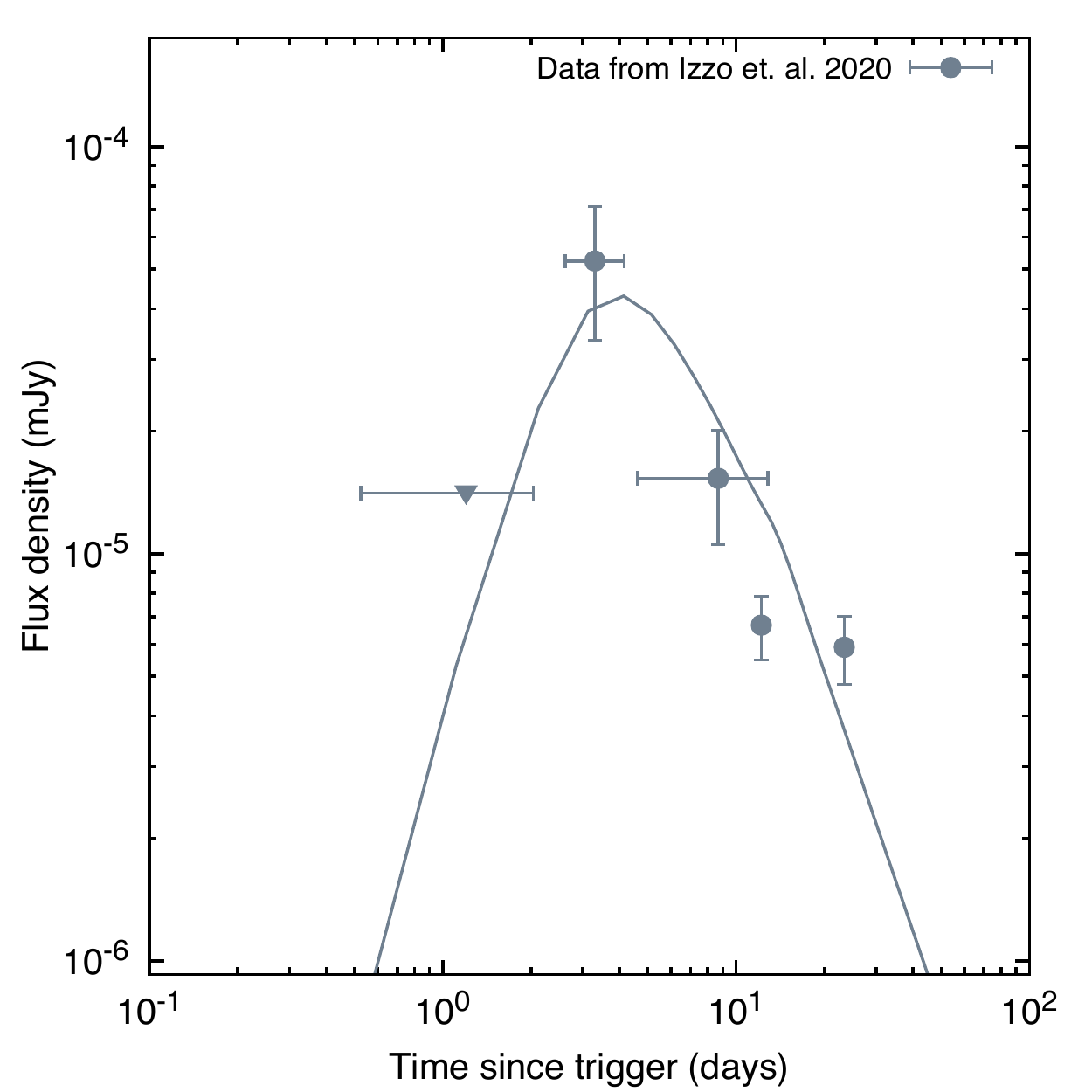}}}
\caption{The X-ray data points of SN 2020bvc with the best-fit curve obtained with the model presented in this article for a stratification parameter of $k=1.5$. The synchrotron light curve are shown at 1 keV. Data points are taken from \cite{2020A&A...639L..11I}.}   
\label{Fit_SN2020bvc}
\end{figure}

\begin{figure}
\vspace{0.4cm}
{\centering
\resizebox*{1.1\textwidth}{0.8\textheight}
{\includegraphics[angle=-90]{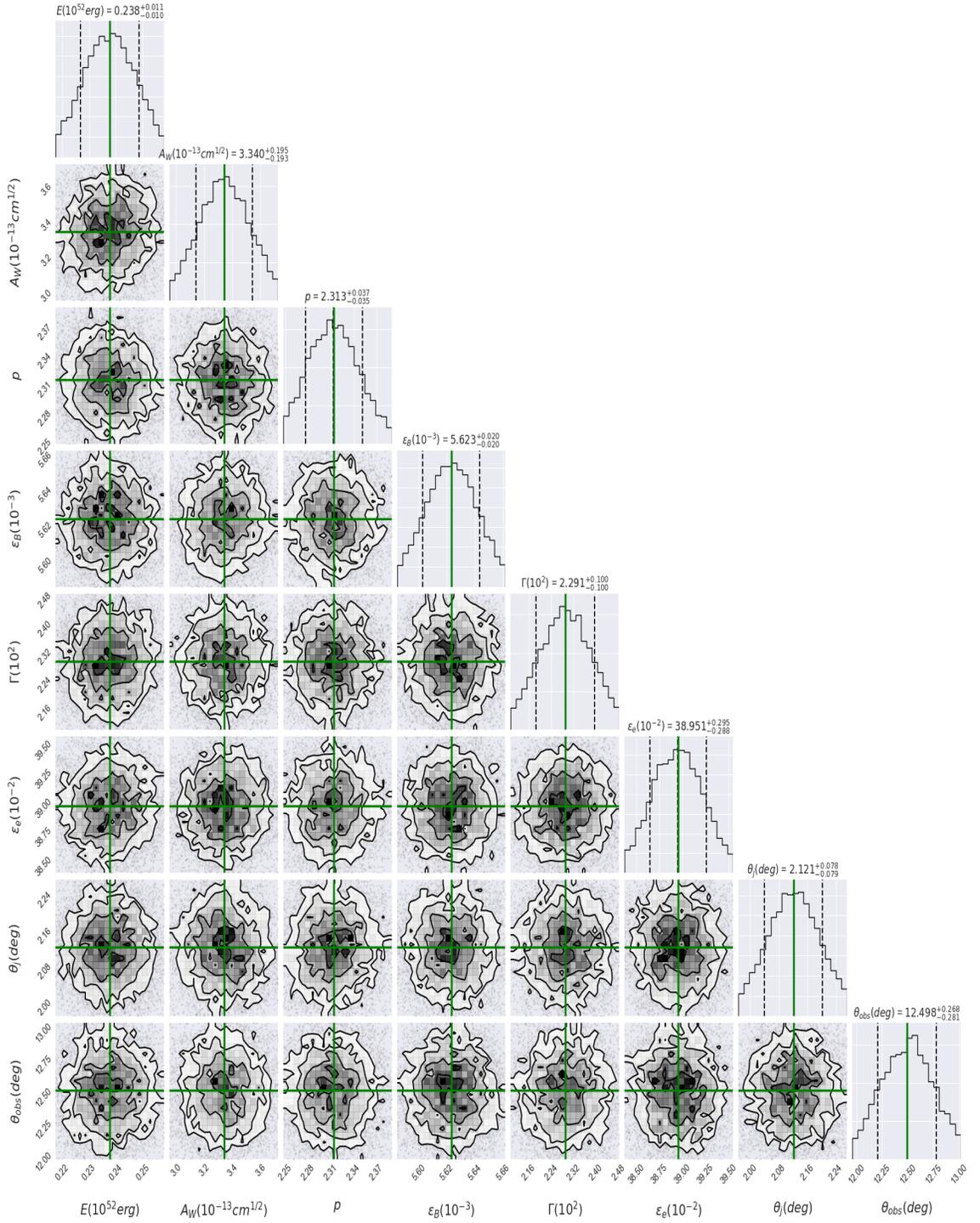}}}
\caption{The same as Figure~GRB 080503, but for SN 2020bvc.}   
\label{SN2020bvc_mcmc}
\end{figure}

We use MCMC simulations to find the best-fit values of the parameters that describe the X-ray afterglow observations. A set of eight parameters, \{$E$, $A_{w}$,  $p$, $\varepsilon_{B}$, $\Gamma$, $\varepsilon_e$, $\theta_{\rm j}$ and $\theta_{\rm obs}$\}, is used to describe the X-ray observations.  To characterize the complete data in this scenario, a total of 17100 samples and 7200 tuning steps are used. Figure~\ref{SN2020bvc_mcmc} exhibits the best-fit values and the median of the posterior distributions of the parameters. In Table~\ref{par_mcmc}, the best-fit values are shown in green, and the medians of the posterior distributions are presented. We assumed a stratified medium with a parameter $k=1.5$, consistent with the proposal by \cite{2020A&A...639L..11I}. Our best fit values, however, are slightly different. We propose that the emission is due to an off-axis jet that is $\approx5$ times more energetic and with half of the off-axis angle compared to the values of \cite{2020A&A...639L..11I}. This discrepancy is due to a different choice of the electron velocity distribution index $p$, as our MCMC simulation suggested $p=2.313^{+0.037}_{-0.035}$, in contrast with the value of $p=2$ used by the cited authors.   Figure \ref{Fit_SN2020bvc} shows the X-ray observations of SN 2020bvc with the best-fit synchrotron light curve generated by the deceleration of an off-axis jet in a medium with stratification parameter $k=1.5$. \\

Our results are consistent with the X-ray observations before and after $\sim 4\,{\rm days}$ since the trigger time; when the observed flux increases and decreases, respectively. Initially, the flux increases with a minimum rise index of $\alpha_{\rm m, ris}>1.65$ and later the observed flux decreases with $\alpha_{\rm dec}=-1.35\pm0.09$ \citep{2020A&A...639L..11I}. The allowed value of the minimum rise index is estimated considering a simple power-law function, the X-ray upper limit and the maximum flux. For instance,  given the best-fit value of $p=2.313^{+0.037}_{-0.035}$ and for $0<k<1.5$, the temporal rise and decay indexes are $1.65\pm 0.03 \leq \alpha_{\rm ris}\leq 4.53\pm0.03$ and $\alpha_{\rm dec}= -1.24\pm 0.03$, respectively, for $\nu_{\rm c}<\nu$ (see Table \ref{Table2}). For ${\rm k>1.5}$, the expected rise index would be $\alpha_{\rm ris} < 1.65$, which cannot reproduce the X-ray observations.

The synchrotron scenario from on-axis outflow in a very-low density environment is  disfavored for $p\sim 2$.  The closure relations of synchrotron on-axis model from an outflow decelerating in a stratified environment for $k$ in general can be estimated.  During the coasting and the deceleration phases the bulk Lorentz factor evolves as $\Gamma\propto t^0$ and $\Gamma\propto t^{\frac{k-3}{8-2k}}$, respectively. Therefore, the synchrotron flux during the slow-cooling regime evolves as $F_{\rm \nu} \propto t^{\frac{12-k(p+5)}{4}}\,\nu^{-\frac{p-1}{2}}$ for  $\nu_{\rm m}<\nu<\nu_{\rm c}$ and $\propto  t^{\frac{8-k(p+2)}{4}}\,\nu^{-\frac{p}{2}}$ for  $\nu_{\rm c} <\nu$ during the coasting phase, and  $F_{\rm \nu} \propto t^{-\frac{12(p-1) + k(5-3p)}{4(4-k)}}\nu^{-\frac{p-1}{2}}$ for $\nu_{\rm m}<\nu<\nu_{\rm c}$ and $\propto t^{-\frac{3p-2}{4}}\nu^{-\frac{p}{2}}$ for $\nu_{\rm c}<\nu$, during the deceleration phase. It is worth noting that for the cooling condition $\nu_{\rm c}<\nu$ and with a value of $p=2.5$,  the temporal evolution is only consistent for $k\lesssim 0.3$

The best-fit values of  bulk Lorentz factor ($\Gamma=2.291^{+0.100}_{-0.100}\times 10^2$), the equivalent-kinetic energy ($E=2.38^{+0.01}_{-0.01}\times 10^{51}\,{\rm erg}$) and the half-opening angle $\theta_{\rm j}= 2.121^{+0.078}_{-0.079}\,{\rm deg}$ indicate that synchrotron emission is produced from a narrowly collimated outflow decelerating in an external medium.  The best-fit values of the viewing angle $\theta_{\rm obs}= 12.498^{+0.268}_{-0.281}\,{\rm deg}$ and the half-opening angle $\theta_{\rm j}= 2.121^{+0.078}_{-0.079}\,{\rm deg}$ are consistent with the lack of early multi-wavelength observations.\\

\begin{figure}
\vspace{0.4cm}
{\centering
\resizebox*{0.9\textwidth}{0.4\textheight}
{\includegraphics{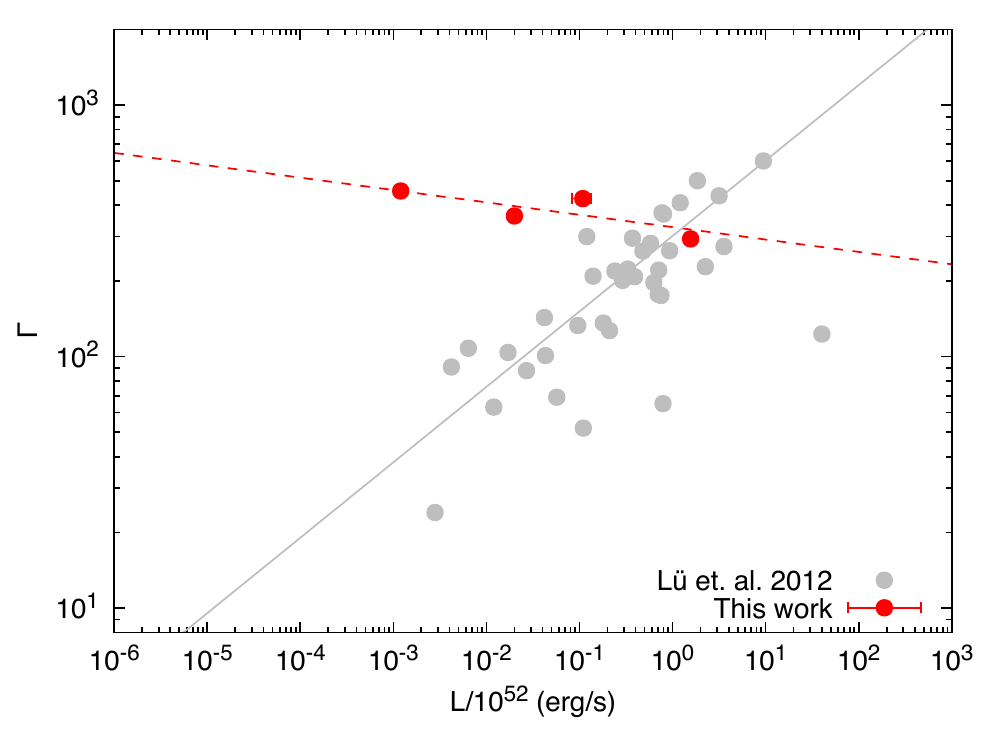}}}
\caption{Diagram of bulk Lorentz factors and Luminosities for the GRBs discussed in    \cite{Fan_2012} and \cite{L__2012},  and for sGRBs described in this work. The red dashed line corresponds to the best-fit $\Gamma=a({\rm L/10^{52}\,erg})^{b}$  with a = $(3.27\pm0.39) \times 10^{2} $ and b = $-(4.9 \pm 0.20) \times 10^{-2}$. }   
\label{gamma-L}
\end{figure}

\cite{L__2012} discovered a correlation between the bulk Lorentz factors and the isotropic gamma-ray luminosities in a sample of GRBs.  \cite{Fan_2012} showed that the correlation of these parameters were consistent with the parameters predicted in the photospheric emission model. Figure \ref{gamma-L} shows the diagram of the bulk Lorentz factors and the isotropic gamma-ray luminosities of  sGRBs described in this work (red) with those (gray) reported in \cite{L__2012}. For off-axis sGRBs, we found an empirical correlation $\Gamma=a({\rm L/10^{52}\,erg})^{b}$  with a = $(3.27\pm0.39) \times 10^{2} $ and b = $-(4.9 \pm 2.0) \times 10^{-2}$.



\vspace{1.5cm}

\section{Constrains on possible afterglow emission}

\subsection{The closest sGRBs detected by Swift ($100 \leq d_z \leq 200$ Mpc)}


\paragraph{GRB 050906}  {\itshape Swift}-BAT was triggered by GRB 050906 at 10:32:18 UTC on September 5, 2005, with a reported location of R.A.=$03^{\textrm{h}}31^{\textrm{m}}13^{\textrm{s}}$ and Dec=$-14^{\circ}37'30''$ (J2000) with a positional accuracy of $3'$\citep{GCN3926}. The light curve revealed an excess in the ($25-100$) keV energy range. The duration and measured fluence in the energy range of 15 - 150~keV were $128\pm16\ \textrm{ms}$ and $(5.9\pm3.2)\times10^{-8}\ \textrm{erg}\ \textrm{cm}^{-2}$, respectively \citep{GCN3935}. \cite{2008MNRAS.384...541D} provided the specifics of Swift's deep optical and infrared observations. According to the authors, no X-ray nor optical/IR afterglow was detected to deep limits, and no residual optical or IR emission was observed.

\paragraph{GRB 070810B}  {\itshape Swift}-BAT was triggered by GRB 070810B at 15:19:17 UTC on August 10, 2007, with a reported location of $\textrm{R.A.}=00^{\textrm{h}}35^{\textrm{m}}46^{\textrm{s}}$ and $\textrm{Dec}=+08^{\circ}50'07''$ (J2000) with a positional accuracy of $3'$\citep{GCN6743}. The KANATA 1.5-m telescope, the Xinlong TNT 80 cm telescope, the 2-m Faulkes Telescope South, the Shajn 2.6 m telescope, and the Keck I telescope (HST) conducted follow-up observations after the initial detection, which are summarized in \cite{2019Preprint}. From the whole observational campaign, only the Shajn telescope detected a source inside the error box of GRB 070810B \citep{GCN6762}.

\paragraph{GRB 080121}  
{\itshape Swift}-BAT was triggered by GRB 080121 at 21:29:55 UTC on January 21, 2008, with a reported location of $\textrm{R.A.}=09^{\textrm{h}}08^{\textrm{m}}56^{\textrm{s}}$, $\textrm{Dec}=+41^{\circ}50'29''$ (J2000) and a positional accuracy of $3'$.
In the ($15-150$) keV energy range, the duration and measured fluence were $(0.7\pm0.2)\, \textrm{s}$ and $(3\pm2)\times10^{-8}\ \textrm{erg}\ \textrm{cm}^{-2}\ \textrm{s}^{-1}$, respectively, according to  \cite{GCN7209}.
Follow-up observations were carried out 2.3 days following the burst using the {\itshape Swift}/UVOT and the  {\itshape Swift}/XRT.
Within the {\itshape Swift}-BAT error circle, however, no X-ray afterglow candidate or sources were discovered \citep{GCN7217,GCN7224}.
Within the {\itshape Swift}-BAT error circle, two galaxies were found, indicating a redshift of $z\sim0.046$ for GRB 080121, however the isotropic energy released would be several orders of magnitude lower than usual short-hard bursts \citep{GCN7210}. 

\paragraph{GRB 100216A}
{\itshape Swift}-BAT and {\itshape Fermi}-GBM were triggered by GRB 100216A at 10:07:00 UTC on February 16, 2010 with a reported location of $\textrm{R.A.}=10^{\textrm{h}}17^{\textrm{m}}03.2^{\textrm{s}}$, $\textrm{Dec}=+35^{\circ}31'27.5''$ (J2000) with a positional accuracy of $3'$.
In the energy range of ($15 - 350$) keV, the duration and measured fluence of the single peak were $0.3\ \textrm{s}$ and $(4.7\pm3)\times10^{-8}\ \textrm{erg}\ \textrm{cm}^{-2}$, respectively \citep{GCN10428}.
The burst was followed up by {\itshape Swift}-XRT and {\itshape Swift}-UVOT from $214.4$ to $249.2\ \textrm{ks}$ after the BAT trigger.
Within the observation, no fading sources were found, but a source presumed to be 1RXS J101702.9+353404 was discovered within the error circle \citep{GCN10435,GCN10442}.

\vspace{1cm}

\begin{figure}
\vspace{0.4cm}
{\centering
\resizebox*{0.9\textwidth}{0.5\textheight}
{\includegraphics{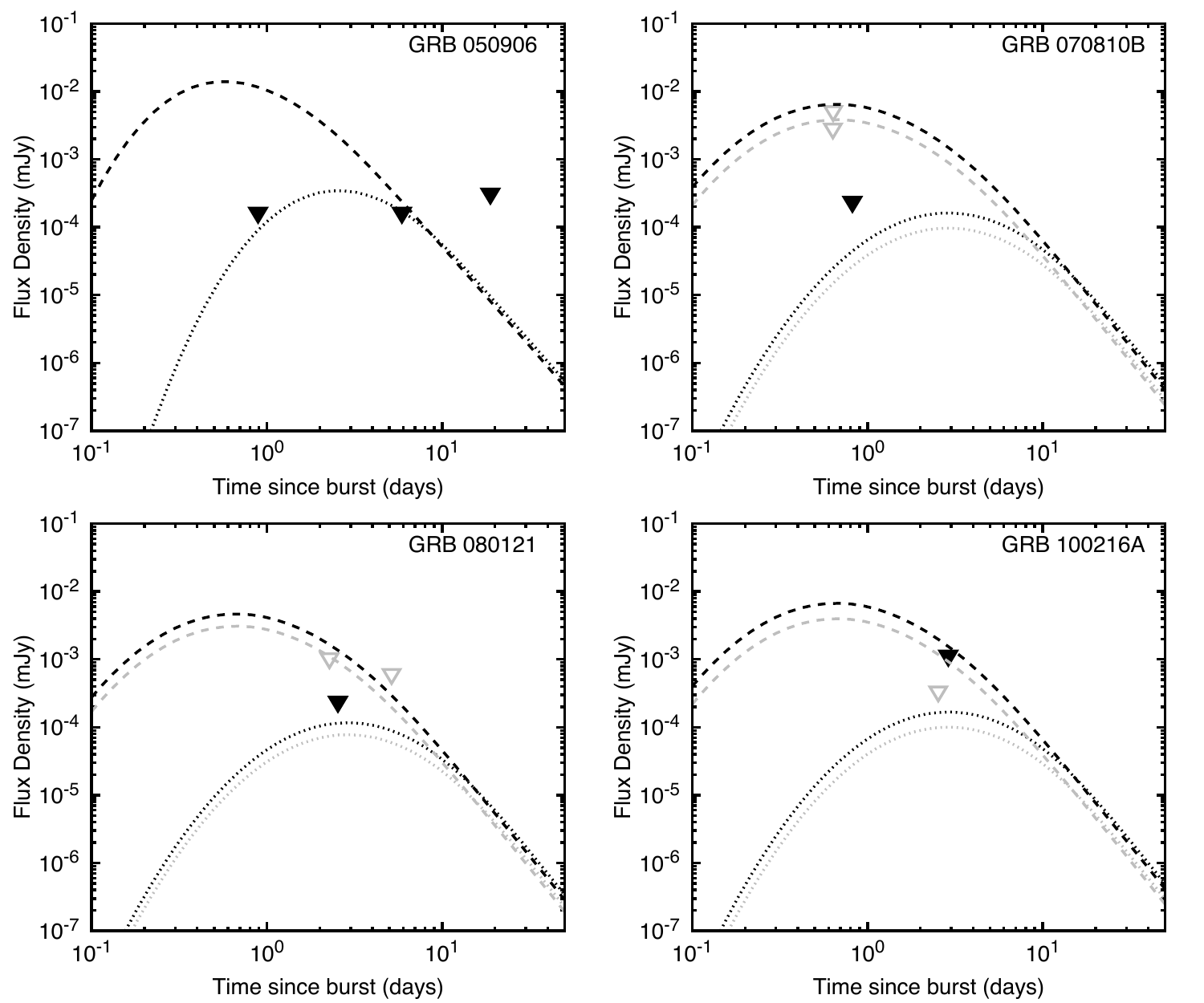}}}
\caption{Optical (u- and r-band) upper limits for GRB 050906, GRB 070810B, GRB 080121 and GRB 100216A with a set of synchrotron light curves evolving in a constant-density medium. The dashed and dotted lines are shown with $n=1\,{\rm cm^{-3}}$ and  $n=10^{-2}\,{\rm cm^{-3}}$, respectively. The parameter values used are $E=5\times 10^{50}\,{\rm erg}$, $\theta_{\rm j}=3\,{\rm deg}$, $\theta_{\rm obs}=15\,{\rm deg}$, $\Gamma=100$, $\varepsilon_{\rm e}=0.3$, $p=2.5$, $\zeta_e=1.0$ and $\varepsilon_{\rm B}=10^{-4}$.  Upper limits are taken from \cite{2020MNRAS.492.5011D}.}   
\label{swift_grbs}
\end{figure}

Figure~\ref{swift_grbs} shows the UV and optical upper limits of the closest sGRBs detected by Swift between 100 and 200 Mpc with the synchrotron light curves from an off-axis outflow decelerating in a constant-density medium. The synchrotron light curves are exhibited at the R-band (black) and the UVW1-band (gray). The parameter values used to generate the synchrotron light curves are reported in Table \ref{TableSwift}. We report two values for each parameter; the upper values correspond to synchrotron light curves on the left-hand panels and lower ones on the right panels.   For typical values of GRB afterglow reported in Table \ref{TableSwift}, the synchrotron emission is ruled out for a density of $n=1\,{\rm cm^{-3}}$, but not for $n=10^{-2}\,{\rm cm^{-3}}$.  The value of the uniform-density medium ruled out in our model is consistent with the mean value reported for sGRBs \citep[e.g., see][]{2014ARA&A..52...43B}. It is worth noting that for values of $\varepsilon_B< 10^{-4.3}$ and $\varepsilon_e< 10^{-1.2}$, the value of the density $n=1\, {\rm cm}^{-3}$ would be allowed.

\subsection{Promising GW events in the third observing run (O3) that could generate electromagnetic emission}


The Advanced LIGO and Advanced Virgo produced 56 non-retracted alerts of gravitational waves candidates during the O3 run, covering almost one year of operations (from 2019 April 01 to 2020 March 27). Nevertheless, three of them have a probability of being terrestrial larger than 50\%.  The O3 observing run was divided into two epochs associated to ``O3a'' (from April 01 to September 30) and ``O3b'' (from November 01 to March 27, 2020).  The GW events in the O3a and O3b runs are listed in the Gravitational Wave Transient (GWTC-2) Catalog 2 \citep{2021PhRvX..11b1053A} and (GWTC-3) Catalog 3  \citep{2021arXiv211103606T}, respectively, where from GCNs there were 31 and 22 candidate events discovered during O3a and O3b respectively. The promising candidates that are consistent with a source with $m_2<3M_\odot$ and that could generate electromagnetic emission are GW190425, GW190426\_152155, GW190814 in GWTC-2 \citep{2021PhRvX..11b1053A}  and GW191219\_163120, GW200105\_162426, GW200115\_042309, GW200210\_092254 in GWTC-3 \citep{2021arXiv211103606T}. Table~\ref{O3} enumerates the main characteristics of these candidates. \\

\begin{figure}
\vspace{0.4cm}
{\centering
\resizebox*{\textwidth}{0.9\textheight}
{\includegraphics{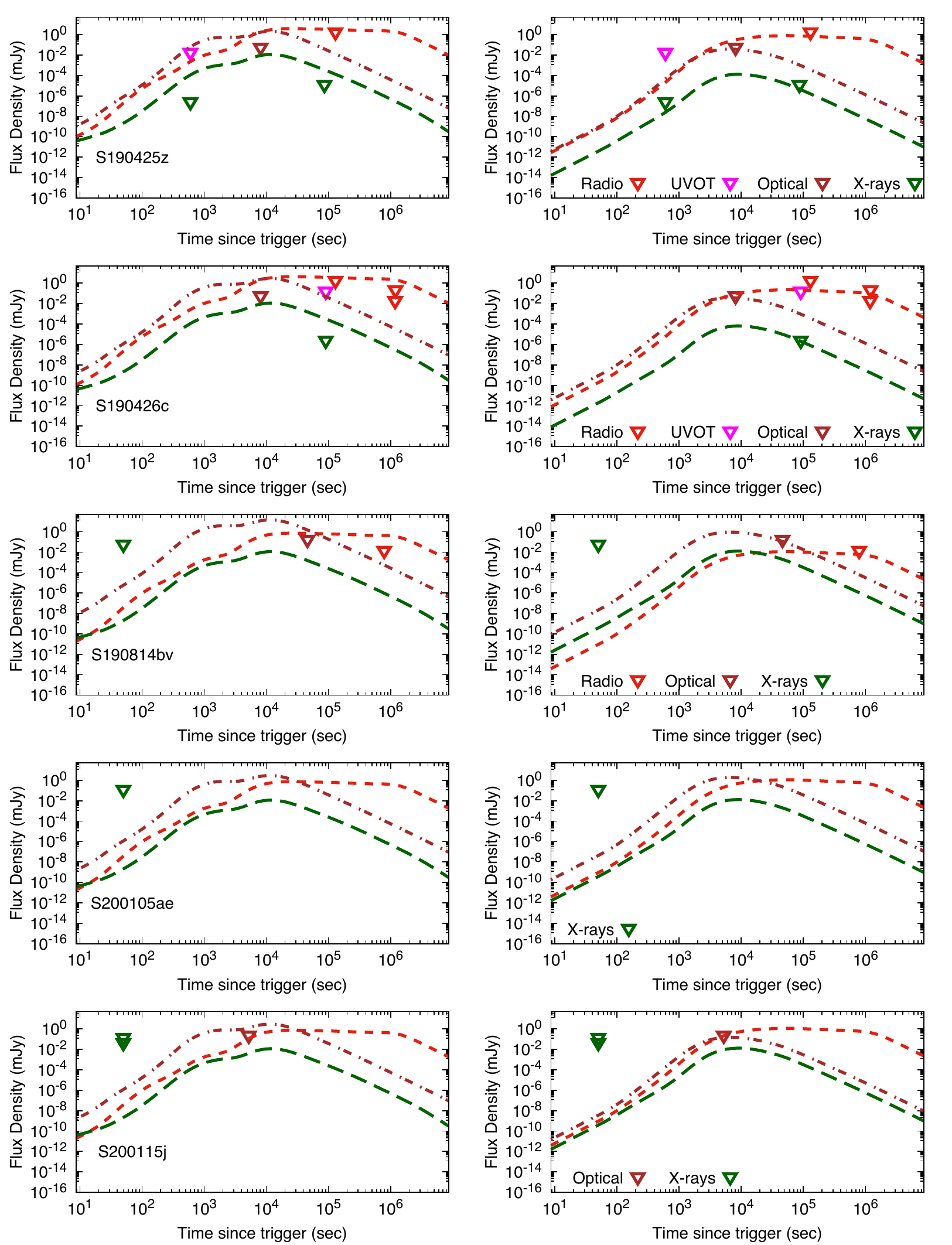}}}
\caption{Promising GW events in the third observing run (O3) that could generate electromagnetic emission  with the synchrotron light curves evolving in a constant-density medium with $n=1\,{\rm cm^{-3}}$ (left panels) and $n=10^{-2}\,{\rm cm^{-3}}$ (right panels), which are decelerating in a constant-density medium. The synchrotron light curves are shown in X-ray (1~keV; green), optical (R-band; brown) and radio (3 GHz; red) bands.  The parameter values used are $E=5\times 10^{50}\,{\rm erg}$, $\theta_{\rm j}=3\,{\rm deg}$, $\theta_{\rm obs}=6\,{\rm deg}$, $\Gamma=100$, $\varepsilon_{\rm e}=0.1$, $p=2.5$, $\zeta_e=1.0$ and $\varepsilon_{\rm B}=10^{-2}$. Upper limits are taken from \cite{}.}   
\label{gw_events}
\end{figure}

Figure~\ref{gw_events} shows the multi-wavelength upper limits of GW events in GWTC-2 and GWTC-3 consistent with a source with $m_2<3M_\odot$ and that could generate electromagnetic emission and the synchrotron light curves from an off-axis outflow decelerating in a constant-density medium with $n=1\,{\rm cm^{-3}}$ (left panels) and $10^{-2}\,{\rm cm^{-3}}$ (right panels).  The synchrotron light curves are presented at 1~keV (green), R-band (brown) (from \cite{2021MNRAS.507.1401B}) and 3 GHz (red).  The parameter values used are $E=5\times 10^{50}\,{\rm erg}$, $\theta_{\rm j}=3\,{\rm deg}$, $\theta_{\rm obs}=6\, {\rm deg}$, $\Gamma=100$, $\varepsilon_{\rm e}=0.1$, $p=2.5$, $\zeta_e=1.0$ and $\varepsilon_{\rm B}=10^{-2}$.  The left-hand panels associated to the S190425z, S190426c, S190814bv and S200115j events show that a uniform-density is ruled out for $n=1\,{\rm cm^{-3}}$, but not for $n=10^{-2}\,{\rm cm^{-3}}$. For instance, we can note that in the panel related to S190426c the synchrotron emission at 1~keV, at the R-band and at 3 GHz is above the upper limits around $\sim$ one day for $n=1\,{\rm cm^{-3}}$, and in the panel associated to S190814bv the synchrotron flux at the R-band and 3 GHz is above the upper limits, but not at 1~keV.  The value of the constant-density medium ruled out in our model is consistent with the densities derived by \cite{2019ApJ...887L..13D,2020arXiv200201950A, 2019ApJ...884L..55G} using distinct off-axis jet models. We need further observations on different timescales and energy bands to derive tighter constraints.


\section{Conclusions}
We have extended the synchrotron off-axis model presented in \cite{2019ApJ...884...71F} initially proposed to describe the multi-wavelength afterglow observations in GRB 170817A. In the current model, we have considered i) the circumburst external medium  as stratified with a profile density $\propto r^{-k}$ with ${\rm k}$ in the range of  $0\leq k \leq 3$, ii) the synchrotron radiation in self-absorption regime, iii) the afterglow emission during the transition from off-axis to on-axis before the lateral expansion phase (relativistic phase), iv) the hydrodynamical evolution computed by numerical simulations and v) a fraction of electrons accelerated by the shock front. The time evolution of the Doppler factor, the half-opening angle, the bulk Lorentz factor, and the deceleration radius computed by numerical simulations are in good agreement with those derived with an analytical approach. The advantage of this general approach (with a density profile $A_{\rm k}\propto r^{-k}$) is that this model allows us to take into account not only both a homogeneous medium ($k=0$) and a wind-like medium ($k=2$) but regions with non-standard stratification parameters, such as $k=1.0$, $1.5$ or $2.5$.\\

We have calculated the synchrotron light curves and presented the closure relations in a stratified environment, including the self-absorption regime for all cooling conditions during the coasting, deceleration (off- and on-axis emission), and the post-jet-break decay phases. We have noted that a break is expected around the transition between fast- and slow-cooling regimes during the deceleration phase when the afterglow emission seems off-axis, but not during other stages. We have analyzed the behavior of the flux for different parameters of the density distribution. We have noticed that the behavior during the relativistic phase approaches flatness as the stratification parameter is raised. On the other hand, we have shown that the time evolution of the light curves after the jet break is independent of $k$, so this model gives freedom to explain the early-time evolution of the radiation while keeping the long-time results invariant. Furthermore,  we have derived the change of the density parameter in the entire phase. We have shown that:
In general, during the coasting and the deceleration phases, when the afterglow emission seems off-axis, a transition between different environments cannot be detected, contrary to the post-jet-break phase and the deceleration phase with on-axis emission.
The synchrotron fluxes do not depend on the density parameter when these are observed at the lowest and the highest frequencies when the afterglow emission seems off- and on-axis, respectively.
During the post-jet-break phase, any transition between a stratified environment and density-constant medium will be better observed in low frequencies, such as radio wavelengths.\\

In particular, we have applied our model to describe the delayed non-thermal emission observed in a sample of bursts with evidence of off-axis emission. In accordance with the best-fit values for sGRBs, we found that i) the constant-density medium required to model the multi-wavelength observations is low  ($10^{-2}\leq n \leq 0.4\,{\rm cm^{-3}}$), indicating that the central engine are located in a low density circumstellar medium, ii) the spectral indexes of the shocked electrons ($2.1\leq p \leq 2.3$) are in the range of those reported after the description of the afterglow observations  \citep[see, e.g.][]{2015PhR...561....1K, 2017ApJ...848...94F,2019ApJ...887..254B, 2019ApJ...872..118B,  2019ApJ...881...12B,2021ApJ...908...39B}, and iii) the half-opening angles ($1.5 \leq \theta_{\rm j}\leq  6.6\,{\rm deg}$),  the bulk Lorentz factors ($130\leq\Gamma\leq 450$) and the equivalent-kinetic energies ($0.2\leq E \leq 5.7 \times 10^{52}\,{\rm erg}$)  provide evidence of narrowly collimated outflow expanding into a constant-density environment. The previous results confirm that the multi-wavelength observations are emitted from the GRB afterglow and indicate a merger of compact objects (two NSs or NS-BH) as possible progenitors of these bursts. The low-density medium agrees with the larger offsets of sGRBs compared with lGRBs. Regarding SN 2020bvc, we found that an atypical stratification parameter of $k=1.5$ is required, supporting the CC-SN scenario. The best-fit values of the half-opening angle $2.121^{+0.078}_{-0.079}\,{\rm deg}$, the viewing angle $12.498^{+0.268}_{-0.281}\,{\rm deg}$, the equivalent-kinetic energy $2.38^{+0.11}_{-0.10}\times 10^{51}\,{\rm erg}$ and the bulk-Lorentz factor $2.291^{+0.100}_{-0.100}\times 10^2$ provide evidence of the scenario of off-axis GRB afterglow.\\ 

We have applied the current model to provide constraints on the possible afterglow emission using multi-wavelength upper limits associated with the closest Swift-detected sGRBs ($<200\ \mathrm{Mpc}$) and the promising GW events. We have shown that the value of the constant-density medium is ruled out, which  is consistent with the mean value of densities reported in for sGRBs \citep[e.g., see][]{2014ARA&A..52...43B} and those derived by S190814bv event  \cite[e.g., see][]{2019ApJ...887L..13D,2020arXiv200201950A, 2019ApJ...884L..55G} using different models.  To derive tighter constraints, further observations on timescales from months to years post-merger phase are required.

\section*{ACKNOWLEDGEMENTS}

We would like to express our gratitude to the anonymous referee for his or her careful reading of the manuscript and insightful recommendations that helped improve the paper's quality and clarity.  We thank Rodolfo Barniol Duran, Tanmoy Laskar, Paz Beniamini and Bing Zhang for useful discussions. NF acknowledges financial  support  from UNAM-DGAPA-PAPIIT  through  grant IN106521. RLB acknowledges support from CONACyT postdoctoral fellowships and the support from the DGAPA/UNAM IG100820 and IN105921.

\bibliography{Bib_150101B}
\addcontentsline{toc}{chapter}{Bibliography}



%
\clearpage
\newpage

\begin{table}
\centering \renewcommand{\arraystretch}{1.85}\addtolength{\tabcolsep}{1.5pt}
\caption{Evolution of the synchrotron light curves ($F_\nu\propto t^{-\alpha}\nu^{-\beta}$) from an off-axis outflow decelerated in a stratified environment}
\label{Table2}

\begin{tabular}{c c c  c c c}
 \hline \hline
&\hspace{0.5cm}     &\hspace{0.5cm}   Coasting phase &\hspace{0.5cm}   Deceleration phase  & \hspace{0.5cm}   Deceleration phase   & \hspace{0.5cm}  Post jet-break phase \\ 
&\hspace{0.5cm}     &\hspace{0.5cm}   &\hspace{0.5cm}   (off-axis afterglow)  & \hspace{0.5cm}    (on-axis afterglow)   & \hspace{0.5cm}     \\ \cdashline{1-6}
 
                     & \hspace{0.5cm}  $\beta $            &   \hspace{0.5cm}  $\alpha $  &  \hspace{0.5cm}  $\alpha $  & \hspace{0.5cm}  $\alpha $ & \hspace{0.5cm} $\alpha $ \\  \hline \hline
$\nu_{\rm a,3} < \nu_{\rm c} < \nu_{\rm m} $ \\ \hline

$\nu < \nu_{\rm a,3} $   	                                 & \hspace{0.5cm} $-2 $                    &\hspace{0.5cm} $-(1+k) $ &\hspace{0.5cm} $-4 $	    &\hspace{0.5cm} $-\frac{4}{4-k} $ &\hspace{0.5cm} $-1 $\\
$ \nu_{\rm a,3} < \nu < \nu_{\rm c} $   	                & \hspace{0.5cm} $-\frac13$           &\hspace{0.5cm} $\frac{6k-11}{3} $	            &\hspace{0.5cm} $\frac{8k-17}{3} $ &\hspace{0.5cm} $\frac{3k-2}{3(4-k)} $ &\hspace{0.5cm} $1 $	\\	
$\nu_{\rm c} < \nu < \nu_{\rm m} $   	                & \hspace{0.5cm} $\frac{1}{2} $    &\hspace{0.5cm} $\frac{3k-8}{4} $	                    &\hspace{0.5cm} $\frac{9k-26}{4} $ &\hspace{0.5cm} $\frac{1}{4} $ &\hspace{0.5cm} $1 $ \\ 	
$\nu_{\rm m} < \nu $   	                                 & \hspace{0.5cm} $\frac{p}{2} $     &\hspace{0.5cm} $\frac{k(p+2)-8}{4} $	                    &\hspace{0.5cm} $\frac{6(p-5)-(p-11)k}{4} $ &\hspace{0.5cm} $\frac{3p-2}{4} $  &\hspace{0.5cm} $p $\\ \hline
$\nu_{\rm a,1} < \nu_{\rm m} < \nu_{\rm c} $ \\\hline

$\nu < \nu_{\rm a,1} $   	                                 & \hspace{0.5cm} $-2 $                    &\hspace{0.5cm} $-2 $	                    &\hspace{0.5cm} $-2 $ &\hspace{0.5cm} $-\frac{2}{4-k} $ &\hspace{0.5cm} $0 $\\
$ \nu_{\rm a,1} < \nu < \nu_{\rm m} $   	        & \hspace{0.5cm} $-\frac{1}{3}$           &\hspace{0.5cm} $\frac{4k-9}{3} $	                    &\hspace{0.5cm} $\frac{8k-21}{3} $ &\hspace{0.5cm} $\frac{k-2}{4-k} $ &\hspace{0.5cm} $\frac{1}{3} $	\\	
$ \nu_{\rm m} < \nu < \nu_{\rm c} $   	                & \hspace{0.5cm} $\frac{p-1}{2} $  &\hspace{0.5cm} $\frac{k(p+5)-12}{4} $	                    &\hspace{0.5cm} $\frac{6(p-5)-(p-11)k}{4} $ &\hspace{0.5cm} $-\frac{12(1-p)+k(3p-5)}{4(4-k)} $ &\hspace{0.5cm} $p $ \\ 	
$\nu_{\rm c} < \nu $   	                                 & \hspace{0.5cm} $\frac{p}{2} $     &\hspace{0.5cm} $\frac{k(p+2)-8}{4} $	                    &\hspace{0.5cm} $\frac{2(3p-16)-(p-10)k}{4} $ &\hspace{0.5cm} $ \frac{3p-2}{4}  $ &\hspace{0.5cm} $p $\\ \hline
$\nu_{\rm m} < \nu_{\rm a,2} < \nu_{\rm c} $ \\\hline 	

$\nu < \nu_{\rm m} $   	                                 & \hspace{0.5cm} $-2 $                    &\hspace{0.5cm} $-2 $	                    &\hspace{0.5cm} $-2 $ &\hspace{0.5cm} $-\frac{2}{4-k} $  &\hspace{0.5cm} $0 $\\
$ \nu_{\rm m} < \nu < \nu_{\rm a,2} $   	        & \hspace{0.5cm} $-\frac52$           &\hspace{0.5cm} $-\frac{8+k}{4} $	                    &\hspace{0.5cm} $\frac{k-14}{4} $ &\hspace{0.5cm} $\frac{3k-20}{4(4-k)} $  &\hspace{0.5cm} $-1 $	\\	
$ \nu_{\rm a,2} < \nu < \nu_{\rm c} $   	                & \hspace{0.5cm} $\frac{p-1}{2} $  &\hspace{0.5cm} $\frac{k(p+5)-12}{4} $	                    &\hspace{0.5cm} $\frac{6(p-5)-(p-11)k}{4} $ &\hspace{0.5cm} $-\frac{12(1-p)+k(3p-5)}{4(4-k)} $ &\hspace{0.5cm} $p $ \\ 	
$\nu_{\rm c} < \nu $   	                                 & \hspace{0.5cm} $\frac{p}{2} $     &\hspace{0.5cm} $\frac{k(p+2)-8}{4} $	                    &\hspace{0.5cm} $\frac{2(3p-16)-(p-10)k}{4} $ &\hspace{0.5cm} $\frac{3p-2}{4} $ &\hspace{0.5cm} $p $\\ \hline

%
%
\hline
\end{tabular}
\end{table}

\begin{table}
\centering \renewcommand{\arraystretch}{1.85}\addtolength{\tabcolsep}{1.5pt}
\caption{Closure relations of synchrotron radiation from off-axis afterglow model in a stratified environment}
\label{Table3}
\begin{tabular}{c c c  c c c}
 \hline \hline
&\hspace{0.5cm}     &\hspace{0.5cm}   Coasting phase &\hspace{0.5cm}   Deceleration phase  & \hspace{0.5cm}   Deceleration phase   & \hspace{0.5cm}  Post jet-break phase \\ 
&\hspace{0.5cm}     &\hspace{0.5cm}   &\hspace{0.5cm}   (off-axis afterglow)  & \hspace{0.5cm}    (on-axis afterglow)   & \hspace{0.5cm}     \\ \cdashline{1-6}

$\nu_{\rm a,3} < \nu_{\rm c} < \nu_{\rm m}$ \\ \hline

$\nu < \nu_{\rm a,3}$   	                                 & \hspace{0.5cm} $-2$                    &\hspace{0.5cm} $\frac{(k+1)\beta}{2}$ &\hspace{0.5cm} $2\beta$	    &\hspace{0.5cm} $\frac{2\beta}{4-k}$ &\hspace{0.5cm} $\frac{\beta}{2}$\\
$ \nu_{\rm a,3} < \nu < \nu_{\rm c}$   	                & \hspace{0.5cm} $-\frac13$           &\hspace{0.5cm} $(11-6k)\beta$	            &\hspace{0.5cm} $(17-8k)\beta$ &\hspace{0.5cm} $\frac{(2-3k)\beta}{4-k}$ &\hspace{0.5cm}  $-3\beta$	\\	
 $\nu_{\rm c} < \nu < \nu_{\rm m}$   	                & \hspace{0.5cm}  $\frac{1}{2}$    &\hspace{0.5cm}  $\frac{(3k-8)\beta}{2}$	                    &\hspace{0.5cm}  $\frac{(9k-26)\beta}{2}$ &\hspace{0.5cm}  $-\frac{\beta}{2}$ &\hspace{0.5cm}  $2\beta$ \\ 	
 $\nu_{\rm m} < \nu$   	                                 & \hspace{0.5cm}  $\frac{p}{2}$     &\hspace{0.5cm}  $\frac{(\beta+1)k-4}{2}$	                    &\hspace{0.5cm}  $\frac{2(6-k)\beta+11k-30}{4}$ &\hspace{0.5cm}  $\frac{3\beta-1}{2}$  &\hspace{0.5cm}  $2\beta$\\ \hline
$\nu_{\rm a,1} < \nu_{\rm m} < \nu_{\rm c}$ \\\hline

 $\nu < \nu_{\rm a,1}$   	                                 & \hspace{0.5cm}  $-2$                    &\hspace{0.5cm}  $\beta$	                    &\hspace{0.5cm}  $\beta$ &\hspace{0.5cm}  $\frac{\beta}{4-k}$ &\hspace{0.5cm}  $0$\\
 $ \nu_{\rm a,1} < \nu < \nu_{\rm m}$   	        & \hspace{0.5cm}  $-\frac{1}{3}$           &\hspace{0.5cm}  $(9-4k)\beta$	                    &\hspace{0.5cm}  $(21-8k)\beta$ &\hspace{0.5cm}  $\frac{3(2-k)\beta}{4-k}$ &\hspace{0.5cm}  $-1$	\\	
 $ \nu_{\rm m} < \nu < \nu_{\rm c}$   	                & \hspace{0.5cm}  $\frac{p-1}{2}$  &\hspace{0.5cm}  $\frac{k(\beta+3)-6}{2}$	                    &\hspace{0.5cm}  $\frac{(6-k)\beta+5k-12}{2}$ &\hspace{0.5cm}  $\frac{3(4-k)\beta+k}{2(4-k)}$ &\hspace{0.5cm}  $2\beta+1$ \\ 	
 $\nu_{\rm c} < \nu$   	                                 & \hspace{0.5cm}  $\frac{p}{2}$     &\hspace{0.5cm}  $\frac{(\beta+1)k-8}{2}$	                    &\hspace{0.5cm}  $\frac{(6-k)\beta+5k-16}{2}$ &\hspace{0.5cm}  $ \frac{3\beta-1}{2} $ &\hspace{0.5cm}  $2\beta$\\ \hline
$\nu_{\rm m} < \nu_{\rm a,2} < \nu_{\rm c}$ \\\hline 	

 $\nu < \nu_{\rm m}$   	                                 & \hspace{0.5cm}  $-2$                    &\hspace{0.5cm}  $\beta$	                    &\hspace{0.5cm}  $\beta$ &\hspace{0.5cm}  $\frac{\beta}{4-k}$  &\hspace{0.5cm}  $0$\\
 $ \nu_{\rm m} < \nu < \nu_{\rm a,2}$   	        & \hspace{0.5cm}  $-\frac52$          &\hspace{0.5cm}  $\frac{(k+8)\beta}{10}$	                    &\hspace{0.5cm}  $\frac{(14-k)\beta}{10}$ &\hspace{0.5cm}  $\frac{(20-3k)\beta}{10(4-k)}$  &\hspace{0.5cm}  $\frac{2\beta}{5}$	\\	
 $ \nu_{\rm a,2} < \nu < \nu_{\rm c}$   	                & \hspace{0.5cm}  $\frac{p-1}{2}$  &\hspace{0.5cm}  $\frac{k(\beta+3)-6}{2}$	                    &\hspace{0.5cm}  $\frac{(6-k)\beta+5k-12}{2}$ &\hspace{0.5cm}  $\frac{3(4-k)\beta+k}{2(4-k)}$ &\hspace{0.5cm}  $2\beta+1$ \\ 	
 $\nu_{\rm c} < \nu$   	                                 & \hspace{0.5cm}  $\frac{p}{2}$     &\hspace{0.5cm}  $\frac{(\beta+1)k-8}{2}$	                    &\hspace{0.5cm}  $\frac{(6-k)\beta+5k-16}{2}$ &\hspace{0.5cm}  $\frac{3\beta-1}{2}$ &\hspace{0.5cm}  $2\beta$\\ \hline

%
%
\hline
\end{tabular}
\end{table}

\clearpage
\begin{table}
\centering \renewcommand{\arraystretch}{1.85}\addtolength{\tabcolsep}{1pt}
\caption{The proportionality constants of the relevant quantities in synchrotron model}
\label{Table1}
\begin{tabular}{l   c  c  c c c}
 \hline \hline
 &  ${\bf k=0}$    &\hspace{0.5cm}   ${\bf k=1.0}$  &\hspace{0.5cm}   ${\bf k=1.5}$  &\hspace{0.5cm}   ${\bf k=2.0}$   &\hspace{0.5cm}   ${\bf k=2.5}$   \\ 
\hline
$A_{\rm k}$ & $1\,{\rm cm^{-3}}$ & $1.4\times 10^{28}\,{\rm cm^{-2}}$ & $2.8\times 10^{36}\,{\rm cm^{-\frac32}}$ & $2.8\times 10^{44}\,{\rm cm^{-1}}$ & $1.4\times 10^{51}\,{\rm cm^{-\frac12}}$ \\
\hline\hline
Coasting phase & & & & &\\
\hline
%
%
$\gamma^0_{\rm m}\,  (\times 10^3)$ & $9.15$ &	$9.15$ &	$9.15$ &	$9.15$ &	$9.15$\\

$\gamma^0_{\rm c}$ & $1.60\times10^5$ &	$1.03\times10$ &	$1.85$ &	$6.65\times10^{-1}$ &	$4.77$\\

$\nu^{\rm 0}_{\rm a,1}\,(\rm Hz)$ & $4.47\times10^{-8}$ &	$1.23$ &	$2.72\times10$ &	$1.72\times10^{2}$ &	$4.96$ \\

$\nu^0_{\rm a,2}\,(\rm Hz)$ & $5.46\times10^{-3}$ &	$2.66$ &	$8.30$ &	$1.64\times10$ &	$4.44$ \\

$\nu^0_{\rm a,3}\,(\times10^{-3}\,\rm Hz)$ & $7.80\times10^{-4}$ &	$1.39$ &	$5.51$ &	$1.25\times10$ &	$2.59$\\

$\nu^{0}_{\rm m}\,(\times10^{4}\,\rm Hz)$ & $1.62\times10^{-2}$ &	$1.74$ &	$4.12$ &	$6.87$ &	$2.57$\\

$\nu^0_{\rm c}\,(\times10^{-3}\,\rm Hz)$ & $4.93\times10^{7}$ &	$2.22\times10$ &	$1.69$ &	$3.63\times10^{-1}$ &	$6.97$ \\

$F^0_{\rm max}\,(\times10^{3}\,{\rm mJy})    $ & $6.57\times10^{-8}$ &	$8.22\times10^{-2}$ &	$1.08$ &	$5.04$ &	$2.62\times10^{-1}$  \\
\hline 
Deceleration phase (Off-axis) & & & & &\\
\hline
%
%
$\gamma^0_{\rm m}\,  (\times 10)$ & $7.56$ &	$3.48$ &	$6.59$ &	$2.06\times10$ &	$1.96\times10^{2}$\\
$\gamma^0_{\rm c}\, (\times 10^{4})$ & $2.80$ &	$4.08\times10^{-1}$ &	$1.98$ &	$8.44\times10$ &	$9.33\times10^{6}$\\
$\nu^{\rm 0}_{\rm a,1}\,(\times 10^{-6}\, \rm Hz)$& $7.55\times10^{-2}$ &	$1.80$ &	$1.35\times10^{-1}$ &	$2.45\times10^{-4}$ &	$6.72\times10^{-13}$ \\
$\nu^0_{\rm a,2}\,(\rm \times10^{-5}\,Hz)$ & $1.00$ &	$1.93$ &	$1.13$ &	$2.47\times10^{-1}$ &	$1.05\times10^{-3}$\\
$\nu^0_{\rm a,3}\,(\rm \times10^{-5}\, Hz)$ & $2.80$ &	$2.11\times10$ &	$4.04$ &	$1.00\times10^{-1}$ &	$3.20\times10^{-6}$\\
$\nu^{0}_{\rm m}\,(\times10^{-6}\,\rm Hz)$ & $3.03$ &	$1.20$ &	$2.58$ &	$7.04$ &	$8.87$\\
$\nu^0_{\rm c}\,(\times10^{-1}\,\rm Hz)$ & $4.17$ &	$1.65\times10^{-1}$ &	$2.32$ &	$1.18\times10^{3}$ &	$2.01\times10^{11}$\\
$F^0_{\rm max}\,(\times10\,{\rm mJy})    $ & $4.10$ &	$7.66$ &	$4.59$ &	$7.02\times10^{-1}$ &	$1.39\times10^{-4}$   \\
\hline 
Deceleration phase (On-axis) & & & & &\\
\hline
%
$\gamma^0_{\rm m}\,  (\times 10^3) $ &   $2.83$ &	$1.43$ &	$1.51$ &	$1.96$ &	$8.19$ \\
$\gamma^0_{\rm c}\,  (\times 10^2) $ & $3.95\times10$ &	$1.36$ &	$1.78$ &	$6.32$ &	$1.28\times10^{4}$\\
$\nu^0_{\rm a,1}\,(\times10^{-3}\,\rm Hz)$ & $1.80\times10^{-2}$ &	$6.77$ &	$4.23$ &	$4.54\times10^{-1}$ &	$1.02\times10^{-6}$ \\
$\nu^0_{\rm a,2}\,(\times10^{-2}\,\rm Hz)$ & $5.74\times10^{-1}$ &	$3.25$ &	$2.83$ &	$1.47$ &	$3.78\times10^{-2}$\\
$\nu^0_{\rm a,3}\,(\times10^{-4}\,\rm Hz)$ & $2.51\times10^{-1}$ &	$6.44$ &	$4.98$ &	$1.47$ &	$1.58\times10^{-3}$\\
$\nu^0_{\rm m}\,(\rm Hz)$ & $3.17$ &	$3.10$ &	$3.11$ &	$3.14$ &	$3.21$ \\
$\nu^0_{\rm c}\,(\times10^{-2}\rm Hz)$ & $6.15\times10^{2}$ &	$2.81$ &	$4.31$ &	$3.28\times10$ &	$7.79\times10^{6}$\\
$F^0_{\rm max}\,(\times10^{3}{\rm mJy})$ & $1.14\times10^{-1}$ &	$1.69$ &	$1.36$ &	$4.95\times10^{-1}$ &	$1.68\times10^{-3}$ \\
\hline \hline
Post-jet-break decay phase & & & & &\\
\hline
%
%
%
$\gamma^0_{\rm m}\,  (\times 10)$ &   $4.68$ &	$1.95$ &	$4.03$ &	$1.69\times10$ &	$1.74\times10^{2}$ \\
$\gamma^0_{\rm  c}\,(\rm \times 10^{4})$ & $3.92$ &	$4.89\times10^{-1}$ &	$2.71$ &	$1.52\times10^{2}$ &	$2.32\times10^{7}$\\
$\nu^{\rm 0}_{\rm a,1}\,(\times 10^{-8}\, \rm Hz)$ & $3.40$ &	$9.82\times10$ &	$6.22$ &	$7.57\times10^{-03}$ &	$1.20\times10^{-11}$\\
$\nu^0_{\rm a,2}\,(\rm \times10^{-6}\,Hz)$ & $4.80$ &	$9.38$ &	$5.43$ &	$1.21$ &	$4.48\times10^{-3}$\\
$\nu^0_{\rm a,3}\,(\rm \times 10^{-5}\, Hz)$ & $2.85$ &	$2.47\times10$ &	$4.19$ &	$6.83\times10^{-2}$ &	$1.60\times10^{-6}$\\
$\nu^{0}_{\rm m}\,(\rm \times 10^{-7}\, Hz)$ & $6.06$ &	$2.10$ &	$5.05$ &	$2.37\times10$ &	$3.13\times10$\\
$\nu^0_{\rm c}\,(\rm \times 10^{-1}\, Hz)$ & $4.25$ &	$1.32\times10^{-1}$ &	$2.29$ &	$1.93\times10^{3}$ &	$5.55\times10^{11}$\\
$F^0_{\rm max}\,({\rm mJy})$ & $2.14\times10$ &	$4.28\times10$ &	$2.41\times10$ &	$4.97$ &	$7.81\times10^{-4}$\\
\hline 

\end{tabular}
\end{table}


\begin{table}
\centering \renewcommand{\arraystretch}{1.85}\addtolength{\tabcolsep}{1.5pt}
\caption{Evolution of the density parameter $F_\nu\propto A_{\rm k}^{\alpha_{\rm k}}$ in each cooling condition of the synchrotron afterglow model} \label{TableDensityParameter}
\begin{tabular}{c c c  c c c}
 \hline \hline
&\hspace{0.5cm}     &\hspace{0.5cm}   Coasting phase &\hspace{0.5cm}   Deceleration phase  & \hspace{0.5cm}   Deceleration phase   & \hspace{0.5cm}  Post jet-break phase \\ 
&\hspace{0.5cm}     &\hspace{0.5cm}   &\hspace{0.5cm}   (off-axis afterglow)  & \hspace{0.5cm}    (on-axis afterglow)   & \hspace{0.5cm}     \\ \cdashline{1-6}

$ $                     & \hspace{0.5cm}  $\beta $            &   \hspace{0.5cm}  $\alpha_{\rm k} $  &  \hspace{0.5cm}  $\alpha_{\rm k} $  & \hspace{0.5cm}  $\alpha_{\rm k}$ & \hspace{0.5cm} $\alpha_{\rm k} $  \\  \hline \hline
$\nu_{\rm a,3} < \nu_{\rm c} < \nu_{\rm m} $ \\ \hline

$\nu < \nu_{\rm a,3} $   	                                 & \hspace{0.5cm} $-2 $                    &\hspace{0.5cm} $-1 $ &\hspace{0.5cm} $0 $	    &\hspace{0.5cm} $-\frac{4}{4-k} $ &\hspace{0.5cm} $-\frac{3}{3-k} $\\
$ \nu_{\rm a,3} < \nu < \nu_{\rm c} $   	                & \hspace{0.5cm} $-\frac13$           &\hspace{0.5cm} $2 $	            &\hspace{0.5cm} $\frac{8}{3} $ &\hspace{0.5cm} $\frac{10}{3(4-k)} $ &\hspace{0.5cm} $-\frac{4}{3(3-k)} $	\\	
$\nu_{\rm c} < \nu < \nu_{\rm m} $   	                & \hspace{0.5cm} $\frac{1}{2} $    &\hspace{0.5cm} $\frac{3}{4} $	                    &\hspace{0.5cm} $\frac{9}{4} $ &\hspace{0.5cm} $0 $ &\hspace{0.5cm} $-\frac{3}{4(3-k)} $ \\ 	
$\nu_{\rm m} < \nu $   	                                 & \hspace{0.5cm} $\frac{p}{2} $     &\hspace{0.5cm} $\frac{p+2}{4} $	                    &\hspace{0.5cm} $\frac{10-p}{4} $ &\hspace{0.5cm} $0 $  &\hspace{0.5cm} $-\frac{p+2}{4(3-k)} $\\ \hline
$\nu_{\rm a,1} < \nu_{\rm m} < \nu_{\rm c} $ \\\hline

$\nu < \nu_{\rm a,1} $   	                                 & \hspace{0.5cm} $-2 $                    &\hspace{0.5cm} $-2 $	                    &\hspace{0.5cm} $0 $ &\hspace{0.5cm} $-\frac{2}{4-k} $ &\hspace{0.5cm} $-\frac{2}{3-k} $\\
$ \nu_{\rm a,1} < \nu < \nu_{\rm m} $   	        & \hspace{0.5cm} $-\frac{1}{3}$           &\hspace{0.5cm} $\frac{4}{3} $	                    &\hspace{0.5cm} $\frac{8}{3} $ &\hspace{0.5cm} $\frac{2}{4-k} $ &\hspace{0.5cm} $-\frac{2}{3(3-k)} $	\\	
$ \nu_{\rm m} < \nu < \nu_{\rm c} $   	                & \hspace{0.5cm} $\frac{p-1}{2} $  &\hspace{0.5cm} $\frac{p+5}{4} $	                    &\hspace{0.5cm} $\frac{11-p}{4} $ &\hspace{0.5cm} $\frac{2}{4-k} $ &\hspace{0.5cm} $\frac{3-p}{4(3-k)} $ \\ 	
$\nu_{\rm c} < \nu $   	                                 & \hspace{0.5cm} $\frac{p}{2} $     &\hspace{0.5cm} $\frac{p+2}{4} $	                    &\hspace{0.5cm} $\frac{10-p}{4} $ &\hspace{0.5cm} $0 $ &\hspace{0.5cm} $-\frac{p+2}{4(3-k)} $\\ \hline
$\nu_{\rm m} < \nu_{\rm a,2} < \nu_{\rm c} $ \\\hline 	

$\nu < \nu_{\rm m} $   	                                 & \hspace{0.5cm} $-2 $                    &\hspace{0.5cm} $0 $	                    &\hspace{0.5cm} $0 $ &\hspace{0.5cm} $-\frac{2}{4-k} $  &\hspace{0.5cm} $-\frac{2}{3-k} $\\
$ \nu_{\rm m} < \nu < \nu_{\rm a,2} $   	        & \hspace{0.5cm} $-\frac52$           &\hspace{0.5cm} $-\frac{1}{4} $	                    &\hspace{0.5cm} $\frac{1}{4} $ &\hspace{0.5cm} $-\frac{2}{4-k} $  &\hspace{0.5cm} $-\frac{7}{4(3-k)} $	\\	
$ \nu_{\rm a,2} < \nu < \nu_{\rm c} $   	                & \hspace{0.5cm} $\frac{p-1}{2} $  &\hspace{0.5cm} $\frac{p+5}{4} $	                    &\hspace{0.5cm} $\frac{11-p}{4} $ &\hspace{0.5cm} $\frac{2}{4-k} $ &\hspace{0.5cm} $\frac{3-p}{4(3-k)} $ \\ 	
$\nu_{\rm c} < \nu $   	                                 & \hspace{0.5cm} $\frac{p}{2} $     &\hspace{0.5cm} $\frac{p+2}{4} $	                    &\hspace{0.5cm} $\frac{10-p}{4} $ &\hspace{0.5cm} $0 $ &\hspace{0.5cm} $-\frac{p+2}{4(3-k)}$\\ \hline

%
%
\hline
\end{tabular}
\end{table}

\clearpage

\begin{table}
\centering \renewcommand{\arraystretch}{2.2}\addtolength{\tabcolsep}{2.4pt}
\caption{Median Values of Parameters used to describe the multi-wavelength afterglow observations of a sample of short and long GRBs}\label{par_mcmc}
\begin{tabular}{ l c c c c c}
\hline
\hline\\ 
{\large   Parameters}	             & 	{\large  GRB 080503}  & {\large GRB 140903A} & {\large  GRB 150101B}& {\large GRB 160821B} & {\large SN 2020bvc} \\

\hline \hline
\\
$E\, (10^{52}\,{\rm erg})$	          &   $2.156^{+0.294}_{-0.295}$  & $3.163^{+0.290}_{-0.296}$   & $1.046^{+0.120}_{-0.124}$   & $0.118^{+0.021}_{-0.021}$  & $0.238^{+0.011}_{-0.010}$\\
${\rm n}\,\, (10^{-2}\,{\rm cm^{-3}})$&   $4.221^{+0.102}_{-0.103}$  & $4.219^{+0.102}_{-0.101}$   & $0.164^{+0.021}_{-0.021}$   & $0.869^{+0.093}_{-0.090}$ & --\\
${\rm \footnote{This value is used when k = 1.5}
A_{\rm w}}\,\, (10^{-13}\,{\rm cm^{1/2}})$&               -- &    --                & --                   & --                  & $3.340^{+0.195}_{-0.193}$\\
${\rm p}\,$	                          &  $2.319^{+0.049}_{-0.049}$   &  $2.073^{+0.048}_{-0.050}$  &  $2.150^{+0.217}_{-0.215}$  &  $2.220^{+0.021}_{-0.021}$ & $2.313^{+0.037}_{-0.035}$ \\
$\varepsilon_{\rm B}\,\,(10^{-3})$    &$3.168^{+0.911}_{-0.897}$     &  $4.050^{+0.909}_{-0.921}$  &  $0.147^{+0.124}_{-0.094}$  &  $0.559^{+0.437}_{-0.365}$ & $5.623^{+0.020}_{-0.020}$ \\
$\varepsilon_{\rm e}\,\,(10^{-2})$	  &$3.745^{+0.298}_{-0.288}$     &  $5.104^{+0.298}_{-0.302}$  &  $1.001^{+0.176}_{-0.164}$ &  $0.155^{+0.021}_{-0.021}$ & $38.951^{+0.295}_{-0.288}$\\
$\Gamma\,(10^2)$	                          &$2.939^{+0.101}_{-0.097}$   &  $3.627^{+0.100}_{-0.100}$ & $4.251^{+0.468}_{-0.453}$ & $4.559^{+0.361}_{-0.358}$ &$2.291^{+0.100}_{-0.100}$\\
$\theta_{\rm j}\,\,(\rm deg)$	      &$6.589^{+0.081}_{-0.078}$     &  $3.210^{+0.080}_{-0.081}$  & $6.887^{+0.662}_{-0.682}$   &  $8.002^{+0.817}_{-0.809}$ & $2.121^{+0.078}_{-0.079}$ \\
$\theta_{\rm obs}\,\,(\rm deg)$	      &$15.412^{+0.268}_{-0.269}$    &  $5.162^{+0.271}_{-0.267}$  & $14.114^{+2.327}_{-2.179}$  &  $10.299^{+1.125}_{-1.135}$ & $12.498^{+0.268}_{-0.281}$ \\
\hline
\end{tabular}
\end{table}

\begin{table}
\centering \renewcommand{\arraystretch}{2.2}\addtolength{\tabcolsep}{2.4pt}
\caption{Values used in the synchrotron light curves of the closest sGRBs detected by the Swift satellite}\label{TableSwift}
\begin{tabular}{ l c c c c}
\hline
\hline\\ 
{\large   Parameters}	                    & 	{\large  GRB 050906}       & {\large GRB 070810B} & 	{\large  GRB 080121}       & {\large GRB 100216A} \\

\hline \hline
\\
$E\, (10^{50}\,{\rm erg})$	&   $5.00$ & $5.00$  & $5.00$ & $5.00$\\
&   $5.00$ & $5.00$  & $5.00$ & $5.00$\\\cdashline{2-5}

$n\, (10^{-2}\,{\rm cm^{-3}})$	&   $1.00$ & $1.00$  & $1.00$ & $1.00$\\
&   $100.00$ & $100.00$  & $100.00$ & $100.00$\\\cdashline{2-5}

$p$	&   $2.50$ & $2.50$  & $2.50$ & $2.50$\\
&   $2.50$ & $2.50$  & $2.50$ & $2.50$\\\cdashline{2-5}

$\varepsilon_B \,\,(10^{-4})$	&   $1.00$ & $1.00$  & $1.00$ & $1.00$\\
&   $1.00$ & $1.00$  & $1.00$ & $1.00$\\\cdashline{2-5}

$\varepsilon_{e}\,\,(10^{-1})$	&   $3.00$ & $3.00$  & $3.00$ & $3.00$\\
&   $3.00$ & $3.00$  & $3.00$ & $3.00$\\\cdashline{2-5}

$\Gamma$	&   $100.0$ & $100.0$  & $150.0$ & $80.0$\\
&   $100.0$ & $100.0$  & $100.0$ & $100.0$\\\cdashline{2-5}

$\theta_{\rm j}\,\,(\rm deg)$	&   $3.00$ & $3.00$  & $3.00$ & $3.00$\\
&   $3.00$ & $3.00$  & $3.00$ & $3.00$\\\cdashline{2-5}

$\theta_{obs}\,\,(\rm deg)$	&   $15.00$ & $15.00$  & $15.00$ & $15.00$\\
&   $15.00$ & $15.00$  & $15.0$ & $15.00$\\\cdashline{2-5}




\hline
\end{tabular}
\end{table}

\begin{table}[]
\centering
\caption{Quantities derived of the promising GW events from  GWTC-2 and GWTC-3}\label{O3}
\begin{tabular}{cllcccl}
\hline \hline
\multirow{2}{*}{} & \multirow{2}{*}{Event} & \multirow{2}{*}{SID}  & \multirow{2}{*}{\begin{tabular}[c]{@{}c@{}}Mass 1\\ ($M_{\odot}$)\end{tabular}} & \multirow{2}{*}{\begin{tabular}[c]{@{}c@{}}Mass 2\\ ($M_{\odot}$)\end{tabular}}  & \multirow{2}{*}{z} & \multirow{2}{*}{References} \\
 &  &  &  &  &  &  \\\\
 \hline 
\parbox{2mm}{\multirow{6}{*}{\rotatebox[origin=c]{90}{GWTC-2}}} & \multirow{2}{*}{GW190425} & \multirow{2}{*}{S190425z} & \multirow{2}{*}{$3.4^{+0.3}_{-0.1}$} & \multirow{2}{*}{$1.44^{+0.02}_{-0.02}$} & \multirow{2}{*}{$0.03^{+0.01}_{-0.02}$} & \multirow{2}{*}{[a], [b], [c]} \\
 &  &  &  &  &  &  \\ 
 & \multirow{2}{*}{GW190426\_152155} & \multirow{2}{*}{S190426c} & \multirow{2}{*}{$5.7^{+3.9}_{-2.3}$} & \multirow{2}{*}{$1.5^{+0.8}_{-0.5}$}  & \multirow{2}{*}{$0.08^{+0.04}_{-0.03}$} & \multirow{2}{*}{[d], [e], [f], [g]} \\
 &  &  &  &  &  &  \\ 
 & \multirow{2}{*}{GW190814} & \multirow{2}{*}{S190814bv} & \multirow{2}{*}{$23.2^{+1.1}_{-1.0}$} & \multirow{2}{*}{$2.59^{+0.08}_{-0.09}$}  & \multirow{2}{*}{$0.05^{+0.009}_{-0.010}$} & \multirow{2}{*}{[h], [i], [l], [m], [n], [o]} \\
 &  &  &  &  &  &   \\ \cline{1-3}
\parbox{2mm}{\multirow{6}{*}{\rotatebox[origin=c]{90}{GWTC-3}}}  & \multirow{2}{*}{GW191219\_163120} & \multirow{2}{*}{} & \multirow{2}{*}{$31.1^{+2.2}_{-2.8}$} & \multirow{2}{*}{$1.17^{+0.07}_{-0.06}$}  & \multirow{2}{*}{$0.11^{+0.05}_{-0.03}$} & \multirow{2}{*}{[p]} \\ 
 &  &  &  &  &  &  \\ 
& \multirow{2}{*}{GW200105\_162426} & \multirow{2}{*}{S200105ae} & \multirow{2}{*}{$9.0^{+1.7}_{-1.7}$} & \multirow{2}{*}{$1.91^{+0.33}_{-0.24}$} & \multirow{2}{*}{$0.06^{+0.02}_{-0.02}$} & \multirow{2}{*}{[q]} \\
 &  &  &  &  &  &  \\ 
 & \multirow{2}{*}{GW200115\_042309} & \multirow{2}{*}{S200115j} & \multirow{2}{*}{$5.9^{+2.0}_{-2.5}$} & \multirow{2}{*}{$1.44^{+0.85}_{-0.29}$} & \multirow{2}{*}{$0.05^{+0.009}_{-0.010}$} & \multirow{2}{*}{[r], [s]} \\
 &  &  &  &  &  &  \\ 
 & \multirow{2}{*}{GW200210\_092254} & \multirow{2}{*}{} & \multirow{2}{*}{$24.1^{+7.5}_{-4.6}$} & \multirow{2}{*}{$2.83^{+0.47}_{-0.42}$} & \multirow{2}{*}{$0.19^{+0.08}_{-0.06}$} & \multirow{2}{*}{[p]} \\ 
 &  &  &  &  &  &  \\  \hline \hline 
\end{tabular}\\
References: \scriptsize{[a]~\cite{UVOT2_S190425z}, [b]~\cite{UVOT_S190425z}, [c]~\cite{LBAND_S190425z}, [d]~\cite{Radio3_S190426c}, [e]~\cite{Radio2_S190426c}, [f]~\cite{Radio_S190426c}, [g]~\cite{UVOT_S190426c},  [h]~\cite{2019GCN.25324....1L}, [i]~\cite{2020ApJ...890..131A}, [l]~\cite{2019ApJ...887L..13D}, [m]~\cite{2020MNRAS.492.5916W}, 12]~\cite{2020arXiv200201950A}, [o]~\cite{2020arXiv200309437V}, [p]~\cite{2021arXiv211103606T}, [q]~\cite{Xrays_S200105ae}, [r]~\cite{BAT_S200115j} and [s]~\cite{XRays_S200115j}}

\end{table}


\clearpage

\newpage
\clearpage
\appendix
\section{Analytical synchrotron afterglow model from an off-axis outflow}\label{Appendix}

\subsubsection{2.1 Coasting phase}

During the coasting phase, the bulk Lorentz factor is constant. The minimum and the cooling electron Lorentz factor are given by

{\small
\bary\label{ele_Lorentz_coast}\nonumber
\gamma_m&=& \gamma^0_m \,g(p)\,\varepsilon_{\rm e,-1}\zeta_{\rm e}^{-1}\Gamma_{2.5},\cr
\gamma_c&=& \gamma^0_c \left(\frac{1+z}{1.022}\right)^{1-k} (1+Y)^{-1}\varepsilon_{B,-4}^{-1}A_{k}^{-1}\,\Gamma_{2.5}^{-1}\,f(\Delta\theta)^{k-1}t^{k-1}_{\rm },\hspace{0.5cm}
\eary
}

where $f(\Delta\theta)\equiv\frac{1}{1-\mu\beta}$.  The synchrotron spectral breaks and maximum flux are 

{\small
\bary\label{ene_break_coast}
\nu_{\rm m}&\simeq& \nu^0_{\rm m}\,\, \left(\frac{1+z}{1.022}\right)^{\frac{k-2}{2}}\,g(p)^2\,\zeta_{\rm e}^{-2} \varepsilon^2_{e,-1}\,\varepsilon_{B,-4}^{\frac12}\,A_{\rm k}^{\frac{1}{2}}\, f(\Delta\theta)^{\frac{2-k}{2}} \Gamma_{2.5}^{2}\,t^{-\frac{k}{2}}_{\rm }\,\cr
\nu_{\rm c}&\simeq& \nu^0_{\rm c} \left(\frac{1+z}{1.022}\right)^{\frac{2-3k}{2}}\, (1+Y)^{-2} \,\varepsilon_{B,-4}^{-\frac32}\,A_{\rm k}^{-\frac{3}{2}}\, f(\Delta\theta)^{\frac{3k-2}{2}} \Gamma_{2.5}^{-2}\,t^{\frac{3k-4}{2}}_{\rm }\,\cr
F_{\rm max} &\simeq& F^0_{\rm max} \left(\frac{1+z}{1.022}\right)^{\frac{3k-2}{2}}\, \zeta_{\rm e} \,\varepsilon_{B,-4}^{\frac12}\,A_{\rm k}^{\frac{3}{2}}\, f(\Delta\theta)^{\frac{12-3k}{2}} d_{z,28.3}^{-2}\Gamma_{2.5}^{-2}\,t^{\frac{6-3k}{2}}_{\rm }\,.
\eary
}

In the self-absorption regime, the spectral breaks are

{\small
\bary\label{SelfAbsorptionCutsCoast}
\nu_{\rm a,1}&\simeq& \nu^0_{\rm a,1}\,g(p)^{-1}\left(\frac{1+z}{1.022}\right)^{\frac{4(k-2)}{5}}\zeta_{e}^{\frac85} A_{k}^{\frac{4}{5}} \varepsilon_{e,-1}^{-1} \varepsilon_{B,-4}^{\frac{1}{5}}\,f(\Delta\theta)^{\frac{4(2-k)}{5}}\Gamma_{2.5}^{-\frac{8}{5}}t^{\frac{3-4k}{5}}\cr
\nu_{\rm a,2}&\simeq& \nu^0_{\rm a,2}\,g(p)^{\frac{2(p-1)}{p+4}} \left(\frac{1+z}{1.022}\right)^{\frac{(k-2)(p+6)}{2(p+4)}}\zeta_{e}^{\frac{2(2-p)}{p+4}} A_{k}^{\frac{p+6}{2(p+4)}} \varepsilon_{B,-4}^{\frac{p+2}{2(p+4)}}\varepsilon_{e,-1}^{\frac{2(p-1)}{p+4}}\,f(\Delta\theta)^{\frac{(2-k)(p+6)}{2(p+4)}}\Gamma_{2.5}^{\frac{2(p-2)}{p+4}}t_{\rm }^{\frac{4-k(p+6)}{2(p+4)}}\cr 
\nu_{\rm a,3}&\simeq& \nu^0_{\rm a,3} \left(\frac{1+z}{1.022}\right)^{\frac{9k-13}{5}}\zeta_{e}^{\frac35}(1+Y)A_{k}^{\frac{9}{5}} \varepsilon_{B,-4}^{\frac{6}{5}}f(\Delta\theta)^{\frac{13-9k}{5}}\Gamma_{2.5}^{\frac{2}{5}} t_{\rm }^{\frac{8-9k}{5}}\,.
\eary
}

Given the spectral breaks and the maximum flux from eqs. \ref{ene_break_coast} and \ref{SelfAbsorptionCutsCoast}, the evolution of synchrotron light curves through the spectral and temporal indexes are listed in Table~\ref{Table2}. Similarly,  the closure relations for each cooling condition during the coasting phase are reported in Table~\ref{Table3}.

\subsubsection{2.2 Deceleration phase (Off-axis afterglow emission)}
During the deceleration phase before afterglow emission enters in the observer's field of view, the bulk Lorentz factor is given by Eq. \ref{Gamma_dec_off}.   The minimum and cooling electron Lorentz factors are  given by
{\small
\bary\label{eLor_syn_ism1}\nonumber
\gamma_m&=& \gamma^0_m\,\left(\frac{1+z}{1.022}\right)^{\frac{3-k}{2}}\zeta_{e}^{-1} A_{k}^{-\frac{1}{2}} \varepsilon_{e,-1}\theta_{j,10}^{-1}\Delta\theta_{15}^{3-k}E_{54}^{\frac{1}{2}} t_{6.7}^{\frac{k-3}{2}}\cr
\gamma_c&=&\gamma^0_c\,\left(\frac{1+z}{1.022}\right)^{-\frac{k+1}{2}} A_{k}^{-\frac{1}{2}} (1+Y)^{-1}\varepsilon_{B,-4}^{-1}  \theta_{j,5}\Delta\theta_{15}^{-(k+1)}E_{54}^{-\frac{1}{2}} t_{6.7}^{\frac{k+1}{2}}\,,
\eary
}
which correspond to a comoving magnetic field given by {\small $B'\propto \,\left(\frac{1+z}{1.022}\right)^{\frac{3}{2}}\varepsilon_{B,-4}^{\frac{1}{2}}\theta_{j,10}^{-1}\Delta\theta_{15}^{3}E_{54}^{\frac{1}{2}} t_{6.7}^{-\frac{3}{2}}$}.
%
%
%
The synchrotron spectral breaks and the maximum flux can be written as
{\small
\bary\label{En_br_syn_off}
\nu_{\rm m}&=& \nu^0_{\rm m}\left(\frac{1+z}{1.022}\right)^{\frac{4-k}{2}}\zeta_{e}^{-2} A_{k}^{-\frac{1}{2}} \varepsilon_{e,-1}^2 \varepsilon_{B,-4}^{\frac{1}{2}}\theta_{j,10}^{-2}\Delta\theta_{15}^{4-k}E_{54} t_{6.7}^{\frac{k-6}{2}}\cr
\nu_{\rm c}&=& \nu^0_{\rm c} \left(\frac{1+z}{1.022}\right)^{-\frac{k+4}{2}} A_{k}^{-\frac{1}{2}} (1+Y)^{-2}\varepsilon_{B,-4}^{-\frac{3}{2}}\theta_{j,10}^{2}\Delta\theta_{15}^{-(k+4)}E_{54}^{-1} t_{6.7}^{\frac{k+2}{2}}\cr 
F_{\rm max} &=& F^0_{\rm max}\left(\frac{1+z}{1.022}\right)^{\frac{5k-8}{2}}\zeta_{e} A_{k}^{\frac{5}{2}} \varepsilon_{B,-4}^{\frac{1}{2}}d_{z,28.3}^{-2}\theta_{j,10}^{2}\Delta\theta_{15}^{5k-18}E_{54}^{-1} t_{6.7}^{\frac{12-5k}{2}}.
\eary
}
In the self-absorption regime, the spectral breaks are given by

{\small
\bary\label{SelfAbsorptionCuts_off}
\nu_{\rm a,1}&\simeq& \nu^0_{\rm a,1}\left(\frac{1+z}{1.022}\right)^{\frac{4(2k-5)}{5}}\zeta_{e}^{\frac85} A_{k}^{\frac{8}{5}} \varepsilon_{e,-1}^{-1} \varepsilon_{B,-4}^{\frac{1}{5}}\theta_{j,10}^{\frac{8}{5}} \Delta\theta_{15}^{\frac{8(2k-5)}{5}}  E_{54}^{-\frac{4}{5}}t_{6.7}^{\frac{15-8k}{5}}\cr
\nu_{\rm a,2}&\simeq& \nu^0_{\rm a,2} \left(\frac{1+z}{1.022}\right)^{-\frac{24-10k-4p+kp}{2(p+4)}}\zeta_{e}^{\frac{2(2-p)}{p+4}} A_{k}^{\frac{10-p}{2(p+4)}} \varepsilon_{B,-4}^{\frac{p+2}{2(p+4)}}\varepsilon_{e,-1}^{\frac{2(p-1)}{p+4}}\theta_{j,10}^{\frac{2(2-p)}{p+4}}\Delta\theta_{15}^{\frac{4(p-6)-k(p-10)}{p+4}}E_{54}^{\frac{p-2}{p+4}}t_{4}^{\frac{16-10k-6p+kp}{2(p+4)}}\cr 
\nu_{\rm a,3}&\simeq& \nu^0_{\rm a,3} \left(\frac{1+z}{1.022}\right)^{\frac{2(4k-5)}{5}}\zeta_{e}^{\frac35}(1+Y)A_{k}^{\frac{8}{5}} \varepsilon_{B,-4}^{\frac{6}{5}}\theta_{j,10}^{-\frac{2}{5}} \Delta\theta_{15}^{\frac{4(4k-5)}{5}} E_{54}^{\frac{1}{5}} t_{6.7}^{\frac{5-8k}{5}}\,.
\eary
}

Given the spectral breaks and the maximum flux from eqs. \ref{En_br_syn_off} and \ref{SelfAbsorptionCuts_off}, the evolution of synchrotron light curves through the spectral and temporal indexes are listed in Table~\ref{Table2}. Similarly,  the closure relations for each cooling condition during the deceleration phase before the afterglow emission is seemed off-axis are reported in Table~\ref{Table3}.

\subsubsection{2.3 Deceleration phase (On-axis afterglow emission)}
As the bulk Lorentz factor becomes $\Gamma\sim\Delta\theta^{-1}$, the afterglow emission becomes on-axis, being in our field of view. For the adiabatic regime, the bulk Lorentz factor is given by Eq. \ref{Gamma_dec_on}.   As $\Delta\theta$ becomes close to zero ($\Delta \theta\approx0$), the bulk Lorentz factor during the relativistic phase $\Gamma \approx 1/2(1-\beta)$, and therefore, the Doppler factor can be approximated to $\delta_D\approx 2\Gamma$.   The minimum and cooling electron Lorentz factors are given by
{\small
\bary\label{eLor_syn_ism2}\nonumber
\gamma_m&=& \gamma^0_m\,\left(\frac{1+z}{1.022}\right)^{\frac{3-k}{2(4-k)}}\zeta_{e}^{-1} A_{k}^{-\frac{1}{2(4-k)}} \varepsilon_{e,-1}E_{54}^{\frac{1}{2(4-k)}} t_{3}^{\frac{k-3}{2(4-k)}}\cr
\gamma_c&=&\gamma^0_c\,\left(\frac{1+z}{1.022}\right)^{-\frac{k+1}{2(4-k)}} A_{k}^{-\frac{5}{2(4-k)}} (1+Y)^{-1}\varepsilon_{B,-4}^{-1} E_{54}^{\frac{2k-3}{2(4-k)}} t_{3}^{\frac{k+1}{2(4-k)}}\,,
\eary
}
which correspond to a comoving magnetic field given by {\small $B'\propto \left(\frac{1+z}{1.022}\right)^{\frac{3}{2(4-k)}}A_{k}^{\frac{3}{2(4-k)}}\varepsilon_{B,-4}^{\frac{1}{2}} E_{54}^{\frac{1-k}{2(4-k)}} t_{3}^{-\frac{3}{2(4-k)}}$}.
%
%
%
The synchrotron spectral breaks and the maximum flux can be written as
{\small
\bary\label{En_br_syn_on}
\nu_{\rm m}&=& \nu^0_{\rm m}\left(\frac{1+z}{1.022}\right)^{\frac{1}{2}}\zeta_{e}^{-2} \varepsilon_{e,-1}^2 \varepsilon_{B,-4}^{\frac{1}{2}} E_{54}^{\frac12} t_{3}^{-\frac{3}{2}}\cr
\nu_{\rm c}&=& \nu^0_{\rm c} \left(\frac{1+z}{1.022}\right)^{-\frac{k+4}{2(4-k)}} A_{k}^{-\frac{4}{4-k}} (1+Y)^{-2}\varepsilon_{B,-4}^{-\frac{3}{2}} E_{54}^{\frac{3k-4}{2(4-k)}} t_{3}^{\frac{3k-4}{2(4-k)}}\cr 
F_{\rm max} &=& F^0_{\rm max}\left(\frac{1+z}{1.022}\right)^{\frac{16-3k}{2(4-k)}}\zeta_{e} A_{k}^{\frac{2}{4-k}} \varepsilon_{B,-4}^{\frac{1}{2}}d_{z,28.3}^{-2} E_{54}^{\frac{8-3k}{2(4-k)}} t_{3}^{-\frac{k}{2(4-k)}}\,.
\eary
}
In the self-absorption regime, the spectral breaks are given by

{\small
\bary\label{SelfAbsorptionCuts_on}
\nu_{\rm a,1}&\simeq& \nu^0_{\rm a,1}\left(\frac{1+z}{1.022}\right)^{\frac{4(2k-5)}{5(4-k)}}\zeta_{e}^{\frac85} A_{k}^{\frac{12}{5(4-k)}} \varepsilon_{e,-1}^{-1} \varepsilon_{B,-4}^{\frac{1}{5}} E_{54}^{-\frac{4(1-k)}{5(4-k)}}t_{3}^{-\frac{3k}{5(4-k)}}\cr
\nu_{\rm a,2}&\simeq& \nu^0_{\rm a,2} \left(\frac{1+z}{1.022}\right)^{\frac{4(p-6)-k(p-10)}{2(4-k)(p+4)}}\zeta_{e}^{\frac{2(2-p)}{p+4}} A_{k}^{\frac{8}{(4-k)(p+4)}} \varepsilon_{B,-4}^{\frac{p+2}{2(p+4)}}\varepsilon_{e,-1}^{\frac{2(p-1)}{p+4}} E_{54}^{\frac{4(p+2)-k(p+6)}{2(4-k)(p+4)}}t_{3}^{-\frac{4(3p+2)+k(2-3p)}{2(4-k)(p+4)}}\cr 
\nu_{\rm a,3}&\simeq& \nu^0_{\rm a,3} \left(\frac{1+z}{1.022}\right)^{\frac{2(4k-5)}{5(4-k)}}\zeta_{e}^{\frac35}(1+Y)A_{k}^{\frac{22}{5(4-k)}} \varepsilon_{B,-4}^{\frac{6}{5}}  E_{54}^{\frac{14-9k}{5(4-k)}} t_{3}^{-\frac{10+3k}{5(4-k)}}\,.
\eary
}

Given the spectral breaks and the maximum flux from eqs. \ref{En_br_syn_on} and \ref{SelfAbsorptionCuts_on}, the evolution of synchrotron light curves through the spectral and temporal indexes are listed in Table~\ref{Table2}. Similarly,  the closure relations for each cooling condition during the deceleration phase once the afterglow emission is observer's field of view are reported in Table~\ref{Table3}.

\subsubsection{2.4 Post-jet-break decay phase}

The bulk Lorentz factor decelerating in the stratified environment far away from the progenitor becomes $\Gamma\sim \theta_{\rm j}^{-1}$ \citep{1999ApJ...519L..17S, 2002ApJ...570L..61G,  2017arXiv171006421G}. During the post-jet-break decay phase, the bulk Lorentz factor evolves following Eq. \ref{Gamma_dec_le}.  The minimum and cooling electron Lorentz factors are  given by
{\small
\bary\label{eLor_syn_ism_l}\nonumber
\gamma_m&=& \gamma^0_m\left(\frac{1+z}{1.022}\right)^{\frac{1}{2}}  \zeta_{e}^{-1} \varepsilon_{\rm e, -1}\, A_{k}^{\frac{1}{2(k-3)}}\,E_{54}^{-\frac{1}{2(k-3)}}\,t_{7}^{-\frac{1}{2}}\cr
\gamma_c&=& \gamma^0_c\left(\frac{1+z}{1.022}\right)^{-\frac{1}{2}}  (1+Y)^{-1}\varepsilon_{B,-4}^{-1}\, A_{\rm k}^{\frac{3}{2(k-3)}}\, E_{54}^{\frac{3-2k}{2(k-3)}}\, t_{7}^{\frac{1}{2}}\,.
\eary
}
In this case, the synchrotron spectral breaks and the maximum flux become
{\small
\bary\label{En_br_syn_br}
\nu_{\rm m}&\simeq& \nu^0_{\rm m} \left(\frac{1+z}{1.022}\right)\zeta_{e}^{-2}A_{\rm k}^{\frac{1}{2(k-3)}}\varepsilon_{e,-1}^2 \varepsilon_{B,-4}^{\frac{1}{2}}\,E_{54}^{\frac{k-4}{2(k-3)}}t_{7}^{-2}\cr
\nu_{\rm c} &\simeq& \nu^0_{\rm c} \left(\frac{1+z}{1.022}\right)^{-1}A_{\rm k}^{\frac{5}{2(k-3)}} \varepsilon_{B,-4}^{-\frac{3}{2}}(1+Y)^{-2} E_{54}^{\frac{4-3k}{2(k-3)}} t_{7}\cr 
F_{\rm max} &\simeq& F^0_{\rm max} \left(\frac{1+z}{1.022}\right)^{3}\zeta_{e}A_{k}^{\frac{1}{2(3-k)}}\varepsilon_{B,-4}^{\frac{1}{2}}d_{z,28.3}^{-2}\,E_{54}^{\frac{8-3k}{2(3-k)}}t_{7}^{-1}\,.
\eary
}
Taking into account the self-absorption regime, the spectral breaks are given by
{\small
\bary\label{SelfAbsorptionCuts_br}
\nu_{\rm a,1}&=& \nu^0_{\rm a,1} \left(\frac{1+z}{1.022}\right)^{-\frac{4}{5}}\zeta_{e}^{\frac85} A_{\rm  k}^{\frac{8}{5(3-k)}} \varepsilon_{e,-1}^{-1} \varepsilon_{B,-4}^{\frac{1}{5}}\,E_{54}^{\frac{4(k-1)}{5(k-3)}}t_{7}^{-\frac{1}{5}}\cr
\nu_{\rm a,2}&=& \nu^0_{\rm a,2} \left(\frac{1+z}{1.022}\right)^{\frac{p-2}{p+4}}\zeta_{e}^{\frac{2(2-p)}{p+4}} A_{\rm k}^{\frac{p-10}{2(p+4)(k-3)}} \varepsilon_{B,-4}^{\frac{p+2}{2(p+4)}}\varepsilon_{e,-1}^{\frac{2(p-1)}{p+4}}\,E_{54}^{\frac{k(p+6)-4(p+2)}{2(p+4)(k-3)}}t_{7}^{-\frac{2(p+1)}{p+4}}\,. 
%
\eary
}

Given the spectral breaks and the maximum flux from eqs. \ref{En_br_syn_br} and \ref{SelfAbsorptionCuts_br}, the evolution of synchrotron light curves through the spectral and temporal indexes are listed in Table~\ref{Table2}. Similarly,  the closure relations for each cooling condition during the post-jet-break decay phase are reported in Table~\ref{Table3}.\\

The terms $\gamma^0_{\rm m}$, $\gamma^0_{\rm c}$, $\nu^0_{\rm m}$, $\nu^0_{\rm c}$, $F^0_{\rm max}$, $\nu^0_{\rm a,1}$, $\nu^0_{\rm a,2}$ and $\nu^0_{\rm a,3}$ during the coasting, deceleration (off- and on-axis emission) and the post-jet-break decay phases are shown in Table~\ref{Table1} for ${\rm k=0,\,1,\,1.5\, {\rm and} \,2}$. 

\end{document}